\documentclass[aps,twocolumn,amsmath,amssymb,floatfix,prl,reprint,showpacs,footinbib,superscriptaddress,longbibliography]{revtex4-1}
\usepackage{amssymb}
\usepackage{amsmath}
\usepackage{dcolumn}
\usepackage{graphicx}
\usepackage{mathrsfs}
\usepackage{subfigure}
\usepackage{booktabs}
\usepackage{color}
\usepackage{docmute}
\usepackage{hyphenat}
\usepackage{lineno}
\usepackage{leftidx}
\usepackage{setspace}

\newcommand{\ket}[1]{\mbox{$|#1\rangle$}}

\usepackage{url}
\usepackage[colorlinks]{hyperref}
\hypersetup{%
    plainpages=true,
    breaklinks=true,
    hypertexnames=false,
    pageanchor=true,
    bookmarksnumbered=true,
    colorlinks=true,
    linkcolor={blue},
    citecolor={red},
    urlcolor={blue},
    anchorcolor={black}
}

\begin{document}

\title{Proposal to Test Quantum Wave-Particle Superposition \\on Massive Mechanical Resonators}

\author{Wei Qin}
\affiliation{Theoretical Quantum Physics Laboratory, RIKEN Cluster for Pioneering Research, Wako-shi, Saitama 351-0198, Japan}
\affiliation{Quantum Physics and Quantum Information Division, Beijing Computational Science Research Center, Beijing 100193, China}

\author{Adam Miranowicz}
\affiliation{Theoretical Quantum Physics Laboratory, RIKEN Cluster for Pioneering Research, Wako-shi, Saitama 351-0198, Japan}
\affiliation{Faculty of Physics, Adam Mickiewicz University,
61-614 Pozna\'n, Poland}

\author{Guilu Long}
\affiliation{State Key Laboratory of Low-Dimensional Quantum Physics and Department of Physics, Tsinghua University, Beijing 100084, China}
\affiliation{Beijing Academy of Quantum Information Science, 100193 Beijing, China}
\affiliation{Beijing National Research Center for Information Science and Technology, 100084 Beijing, China}

\author{J. Q. You}
\email{jqyou@zju.edu.cn}
\affiliation{Department of Physics and State Key Laboratory of Modern Optical Instrumentation, Zhejiang University, Hangzhou 310027, China}
\affiliation{Quantum Physics and Quantum Information Division, Beijing Computational Science Research Center, Beijing 100193, China}

\author{Franco Nori}
\email{fnori@riken.jp}
\affiliation{Theoretical Quantum Physics Laboratory, RIKEN Cluster for Pioneering Research, Wako-shi, Saitama 351-0198, Japan}
\affiliation{Department of Physics, The University of Michigan, Ann Arbor, Michigan 48109-1040, USA}

\begin{abstract}
We present and analyze a proposal for a macroscopic quantum delayed-choice experiment with massive mechanical
resonators. In our approach, the electronic spin of a single
nitrogen-vacancy impurity is employed to control the coherent
coupling between the mechanical modes of two carbon nanotubes. We
demonstrate that a mechanical phonon can be in a coherent
superposition of wave and particle, thus exhibiting both behaviors
at the same time. We also discuss the mechanical noise
tolerable in our proposal and predict a critical temperature below
which the morphing between wave and particle states can be
effectively observed in the presence of environment-induced
fluctuations. Furthermore, we describe how to amplify
single-phonon excitations of the mechanical-resonator
superposition states to a macroscopic level, via squeezing the mechanical modes. This approach corresponds to the phase-covariant cloning. Therefore,
our proposal can serve as a test of macroscopic quantum
superpositions of massive objects even with large excitations. This work, which describes a fundamental test of the limits of
quantum mechanics at the macroscopic scale, would have implications for
quantum metrology and quantum information processing.

\end{abstract}

\maketitle
\section{introduction}
Wave-particle duality lies at the heart of quantum physics. According to Bohr's complementarity principle~\cite{bohr1984quantum}, a quantum system may behave either as a wave or as a particle depending on the measurement apparatus, and both behaviors are never observed simultaneously. This can be well demonstrated via a single photon Mach-Zehnder interferometer, as depicted in Fig. \ref{figschematic}(a). An incident photon is split, at an input beam splitter BS$_{1}$, into an equal superposition of being in the upper and lower paths. This is followed by a phase shift $\phi$ in the upper path. At the output beam splitter BS$_{2}$, the paths are recombined and the detection probability in the detector D$_{1}$ or D$_{2}$ depends on the phase $\phi$, heralding the wave nature of a single photon. If, however, BS$_{2}$ is absent, the photon is detected with probability $1/2$ in each detector, and thus, shows its particle nature. In Wheeler's delayed-choice experiment~\cite{wheeler1978mathematical,ma2016delayed}, the decision of whether or not to insert BS$_{2}$ is randomly made after a photon is already inside the interferometer. The arrangement rules out a hidden-variable theory, which suggests that the photon may determine, in advance, which behavior, wave or particle, to exhibit through a hidden variable~\cite{hellmuth1987delayed,kim2000delayed,jacques2007experimental,
jacques2008delayed,manning2015wheeler,liu2017information,vedovato2017extending,chaves2018causal}. Recently, a quantum delayed-choice experiment, where BS$_{2}$ is engineered to be in a quantum superposition of being present and absent, has been proposed~\cite{ionicioiu2011proposal}. Such a version allows a single system to be in a quantum superposition of a wave and a particle, so that both behaviors can be observed in a single measurement apparatus at the same time~\cite{adesso2012quantum, shadbolt2014testing}. This extends the conventional boundary of Bohr's complementarity principle. The quantum delayed-choice experiment has already been implemented in nuclear magnetic resonance~\cite{roy2012nmr,auccaise2012experimental,xin2015realization}, optics~\cite{tang2012realization,peruzzo2012quantum,kaiser2012entanglement,yan2015experimental,rab2017entanglement,long2018realistic}, and superconducting circuits~\cite{zheng2015quantum,liu2017twofold}. However, all these experiments were performed essentially at the \emph{microscopic} scale.

Here, as a step in the \emph{macroscopic} test for a coherent wave-particle superposition on massive objects, we propose and analyze an approach for a mechanical quantum delayed-choice experiment. Mechanical systems are not only being explored now for potential quantum technologies~\cite{blencowe2004quantum,aspelmeyer2014cavity}, but they also have been considered as a promising candidate to test fundamental principles in quantum theory~\cite{poot2012mechanical}.  
In this manuscript, we demonstrate that, similar to a single photon, the mechanical phonon can be prepared in a quantum superposition of both a wave and a particle. The basic idea is to use a single nitrogen-vacancy (NV) center in diamond to control the coherent coupling between two separated carbon nanotubes (CNTs)~\cite{iijima1991helical,liu2019sensing}. We focus on the electronic ground state of the NV center, which is a spin $S=1$ triplet with a zero-field splitting $D\simeq2\pi\times2.87$~GHz between spin states $|0\rangle$ and $|\pm1\rangle$ [see Fig.~\ref{figschematic}(b)]. If the spin is in $|0\rangle$, the mechanical modes are decoupled, and otherwise are coupled. Moreover, the mechanical noise tolerated by our proposal is evaluated and we show a critical temperature, below which the coherent signal is resolved.

\section{results}
\begin{figure}[t]
\centering
\includegraphics[width=8.0cm]{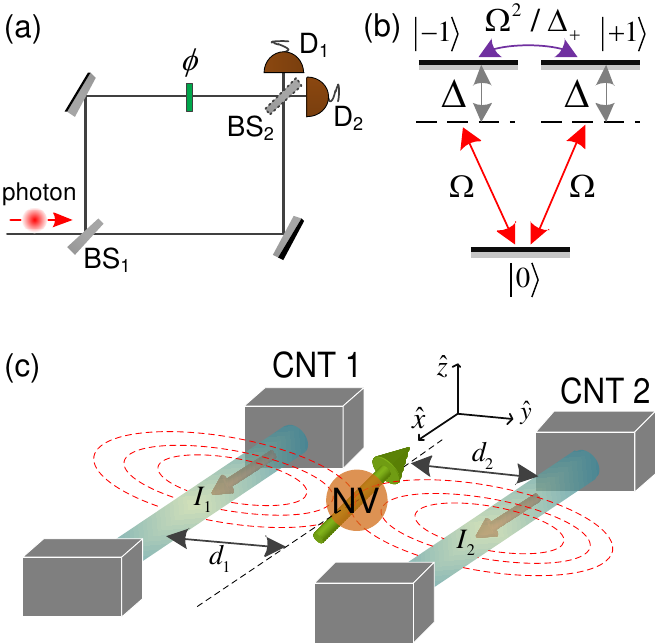}
\caption{(a) Demonstration of the wave-particle duality using a Mach-Zehnder interferometer. A single photon is first split at the input beam splitter BS$_{1}$, then undergoes a phase shift $\phi$ and finally is observed at detectors D$_{1}$ and D$_{2}$. The photon behaves as a wave if the output beam splitter BS$_{2}$ is inserted, or as a particle if BS$_{2}$ is removed. In quantum delayed-choice experiments, BS$_{2}$ is set in a quantum superposition of being present and absent, and consequently, the photon can simultaneously exhibit its wave and particle nature. (b) Level structure of the driven NV spin in the electronic ground state. Here we have assumed that the Zeeman splitting between the spin states $|\pm1\rangle$ is eliminated by applying an external field. (c) Schematic representation of a mechanical quantum delayed-choice experiment with an NV electronic spin and two CNTs. The mechanical vibrations of the CNTs are completely decoupled or coherently coupled, depending, respectively, on whether or not the intermediate spin is in the spin state $|0\rangle$, with the dc current $I_{k}$ through the $k$th CNT, and the distance $d_{k}$ between the spin and the $k$th CNT.}\label{figschematic}
\end{figure}
\emph{Physical model.}---We consider a hybrid system~\cite{Buluta2011,xiang2013hybrid} consisting of two (labelled as $k=1,2$) parallel CNTs and an NV electronic spin, as illustrated in Fig.~\ref{figschematic}(c). The CNTs, both suspended along the $\hat{x}$-direction, carry dc currents $I_{1}$ and $I_{2}$, respectively, while the spin is placed between them, at a distance $d_{1}$ from the first CNT and at a distance $d_{2}$ from the second CNT. When vibrating along the $\hat{y}$-direction, the CNTs can parametrically modulate the Zeeman splitting of the intermediate spin through the magnetic field, yielding a magnetic coupling to the spin~\cite{rabl2009strong,rabl2010quantum,kolkowitz2012coherent,li2016hybrid,cao2017entangling}. For simplicity, below we assume that the CNTs are identical such that they have the same vibrational frequency $\omega_{m}$ and the same vibrational mass $m$. The mechanical vibrations are modelled by quantized harmonic oscillators with a Hamiltonian
\begin{equation}
H_{\text{mv}}=\sum_{k=1,2}\hbar\omega_{m}b_{k}^{\dag}b_{k},
\end{equation}
where $b_{k}$ ($b_{k}^{\dag}$) denotes the phonon annihilation (creation) operator. The Hamiltonian characterizing the coupling of the mechanical modes to the spin is
\begin{equation}
H_{\text{int}}=\sum_{k=1,2}\hbar g_{k}S_{z}q_{k},
\end{equation}
where $S_{z}=|+1\rangle\langle+1|-|-1\rangle\langle-1|$ is the $z$-component of the spin, $q_{k}=b_{k}+b_{k}^{\dag}$ represents the canonical phonon position operator, and $g_{k}=\mu_{B}g_{s}y_{\text{zp}}G_{k}/\hbar$ refers to the Zeeman shift corresponding to the zero-point motion $y_{\text{zp}}=\left[\hbar/\left(2m\omega_{m}\right)\right]^{1/2}$. Here, $\mu_{B}$ is the Bohr magneton, $g_{s}\simeq2$ is the Land\'{e} factor, and $G_{k}=\mu_{0}I_{k}/\left(2\pi d_{k}^{2}\right)$ is the magnetic-field gradient, where $\mu_{0}$ is the vacuum permeability. In order to mediate the coherent coupling of the CNT mechanical modes through the spin, we apply a time-dependent magnetic field
\begin{equation}
B_{x}\left(t\right)=B_{0}\cos\left(\omega_{0}t\right),
\end{equation}
with amplitude $B_{0}$ and frequency $\omega_{0}$, along the $\hat{x}$-direction, to drive the $|0\rangle\rightarrow|\pm1\rangle$ transitions with Rabi frequency \begin{equation}
\Omega=\frac{\mu_{B}g_{s}B_{0}}{2\sqrt{2}\hbar}.
\end{equation}
We apply a static magnetic field
\begin{equation}
B_{z}=\sum_{k=1,2}\left(-1\right)^{k}d_{k}G_{k},
\end{equation}
along the $\hat{z}$-direction to eliminate the Zeeman splitting between the spin states $|\pm1\rangle$~\cite{li2016hybrid}. This causes the same Zeeman shift,
\begin{equation}
\Delta=\Delta_{-}+\frac{3\Omega^{2}}{\Delta_{+}},
\end{equation}
where $\Delta_{\pm}=D\pm\omega_{0}$, to be imprinted on $|\pm1\rangle$, and a coherent coupling, of strength $\Omega^{2}/\Delta_{+}$, between them, as shown in Fig.~\ref{figschematic}(b). We can, thus, introduce a dark state
\begin{equation}
|D\rangle=\left(|+1\rangle-|-1\rangle\right)/\sqrt{2},
\end{equation}
and a bright state
\begin{equation}
|B\rangle=\left(|+1\rangle+|-1\rangle\right)/\sqrt{2},
\end{equation}
with an energy splitting $\simeq2\Omega^{2}/\Delta_{+}$. In this case, the spin state $|0\rangle$ is decoupled from the dark state, and is dressed by the bright state. Under the assumption of $\Omega/\Delta\ll1$, the dressing will only increase the energy splitting between the dark and bright states to
\begin{equation}
\omega_{q}\simeq2\Omega^{2}\left(\frac{1}{\Delta}+\frac{1}{\Delta_{+}}\right).
\end{equation}
This yields a spin qubit with $|D\rangle$ as the ground state and $|B\rangle$ as the exited state. The spin-CNT coupling Hamiltonian is accordingly transformed to
\begin{equation}
H_{\text{int}}\simeq\sum_{k=1,2}\hbar g_{k}\sigma_{x}q_{k},
\end{equation}
where $\sigma_{x}=\sigma_{+}+\sigma_{-}$, with $\sigma_{-}=|D\rangle\langle B|$ and $\sigma_{+}=\sigma_{-}^{\dag}$. When we further restrict our discussion to a dispersive regime $\omega_{q}\pm\omega_{m}\gg |g_{k}|$, the spin qubit becomes a quantum data bus, allowing for mechanical excitations to be exchanged between the CNTs. By using a time-averaging \nohyphens{treatment}~\cite{gamel2010time,qin2018exponentially}, the unitary dynamics of the system is then described by an effective Hamiltonian (see Supplementary Section I for a detailed derivation), $H_{\text{eff}}=H_{\text{cnt}}\otimes\sigma_{z}$, where
\begin{align}\label{eq:effective-Hamiltonian}
H_{\text{cnt}}=&\frac{2\hbar\omega_{q}}{\omega_{q}^{2}-\omega_{m}^{2}}\left[\sum_{k=1,2}
g_{k}^{2}b_{k}^{\dag}b_{k}+g_{1}g_{2}\left(b_{1}b_{2}^{\dag}+{\rm H.c.}\right)\right],
\end{align}
and $\sigma_{z}=|B\rangle\langle B|-|D\rangle\langle D|$. The Hamiltonian $H_{\text{cnt}}$ includes a coherent spin-mediated CNT-CNT coupling in the beam-splitter form, which is conditioned on the spin state. Here, we neglect the direct CNT-CNT coupling much smaller than the spin-mediated coupling, as is described in Supplementary Section I. Furthermore, we find that the decoupling of one CNT from the spin gives rise to a spin-induced shift of the vibrational resonance of the other CNT. Hence, the dynamics described by $H_{\text{eff}}$ can be used to implement controlled Hadamard and phase gates.

\emph{Quantum delayed-choice experiment with mechanical resonators.}---Let us first discuss the Hadamard gate. Having $I_{k}=I$ and $d_{k}=d$ gives a symmetric coupling $g_{k}=g$, and a mechanical beam-splitter coupling of strength
\begin{equation}
J=\frac{2g^{2}\omega_{q}}{\omega^{2}_{q}-\omega_{m}^{2}}.
\end{equation}
Unitary evolution for a time $\tau_{0}=\pi/\left(4J\right)$ then leads to
\begin{align}
b_{1}\left(\tau_{0}\right)=&\left(b_{1}-ib_{2}\right)/\sqrt{2},\\
b_{2}\left(\tau_{0}\right)=&\left(b_{2}-ib_{1}\right)/\sqrt{2}.
\end{align}
For the phase gate, we can turn off the current, for example, of the second CNT, so that $g_{1}=g$ and $g_{2}=0$. In this case, a dispersive shift of $\simeq J$ is imprinted into the vibrational resonance of the first CNT, which in turn introduces a relative phase $\phi\simeq J\tau_{1}$ after a time $\tau_{1}$ under unitary evolution. Note that, here, both Hadamard and phase gates are controlled operations conditional on the spin state, as mentioned before. The two gates and their timing errors are analyzed in detail in the Supplementary Section II.  

\begin{figure}[t]
\centering
\includegraphics[width=7.5cm]{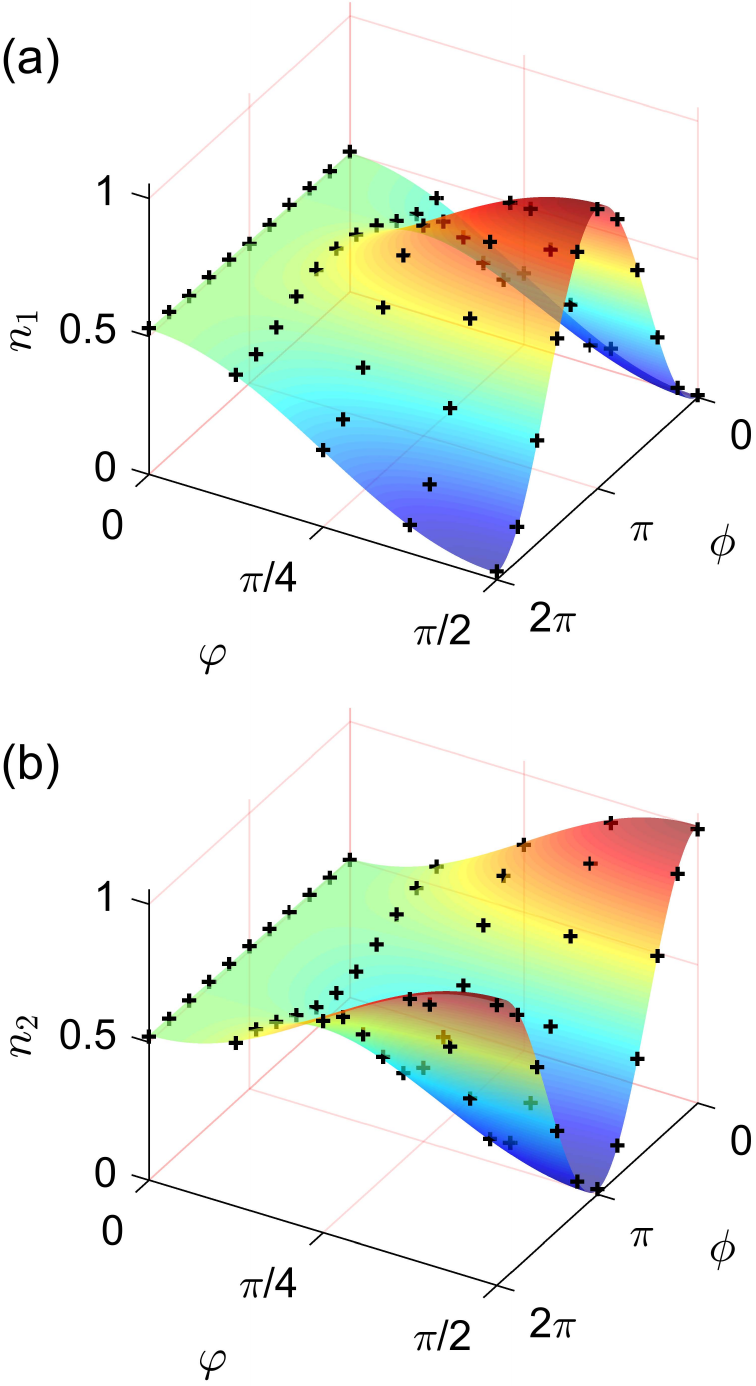}
\caption{Morphing between particle and wave characteristics of a
	CNT mechanical phonon. Phonon occupation (a) $n_{1}$ and (b)
	$n_{2}$ as a function of the relative phase $\phi$ and the
	rotation angle $\varphi$. The analytical results (colored
	surfaces) are in excellent agreement with the numerical
	simulations (black symbols). Here, in addition to
	$\gamma_{s}/2\pi=200\gamma_{m}/2\pi=80$~Hz, we assume that
	$g/2\pi=100$~kHz, $\omega_{m}/2\pi=2$~MHz, $\Omega=10\omega_{m}$,
	and $\Delta_{-}=142\omega_{m}$, resulting in
	$\omega_{q}\simeq1.5\omega_{m}$ and then
	$J/2\pi\simeq\times12$~kHz, and that $n_{\text{th}}=100$,
	corresponding to an environmental temperature of $\simeq
	10$~mK.}\label{fig-morphing}
\end{figure}

We now turn to the quantum delayed-choice experiment with the macroscopic CNTs. We assume that the hybrid system is initially prepared in the state
\begin{equation}
|\Psi\rangle_{i}=\left(b_{1}^{\dag}\otimes \mathcal{I}_{2}|\text{vac}\rangle\right)\otimes|D\rangle,
\end{equation}
where $|\text{vac}\rangle$ refers to the phonon vacuum and $\mathcal{I}_{k}$ is the identity operator for the $k$-th CNT. After the initialization, the currents are tuned to be $I_{k}=I$, to drive the system for a time $\tau_{0}$, and the resulting Hadamard operation splits the single phonon into an equal superposition across both CNTs. Then, we turn off $I_{2}$ for a time $\tau_{1}$ to accumulate a relative phase between the CNTs. While achieving the desired phase $\phi$, we turn on $I_{2}$ following a spin single-qubit rotation $|D\rangle\rightarrow\cos\left(\varphi\right)|0\rangle+\sin\left(\varphi\right)|D\rangle$~\cite{huang2011observation,
lillie2017environmentally,xing2017experimental} with $\varphi$ a rotation angle, and hold for another $\tau_{0}$ for a Hadamard operation. Therefore, this Hadamard gate is in a quantum superposition of being both present and absent. The three steps correspond, respectively, to the input beam splitter, the phase shifter and the quantum output beam splitter acting in sequence on a single photon in the Mach-Zehnder interferometer, as shown in Fig.~\ref{figschematic}(a). The final state of the system therefore becomes
\begin{equation}\label{eq:finial-state}
|\Psi\rangle_{f}=\cos\left(\varphi\right)|\text{particle}\rangle|0\rangle+\sin\left(\varphi\right)|\text{wave}\rangle|D\rangle,
\end{equation}
where
\begin{align}\label{eq:final-state-particle}
|\text{particle}\rangle=&\frac{1}{\sqrt{2}}\left[\exp\left(i\phi\right)b_{1}^{\dag}+ib_{2}^{\dag}\right]
|\text{vac}\rangle,\\
\label{eq:final-state-wave}
|\text{wave}\rangle=&\frac{1}{2}\left\{\left[\exp\left(i\phi\right)-1\right]b_{1}^{\dag}
+i\left[\exp\left(i\phi\right)+1\right]b_{2}^{\dag}\right\}|\text{vac}\rangle,
\end{align}
describe the particle and wave behaviors, respectively. The coherent evolution of the system is given in more detail in Supplementary Section II. We find from Eq.~(\ref{eq:finial-state}) that the mechanical phonon is in a quantum superposition of both a wave and a particle, and thus can exhibit both characteristics simultaneously. By applying microwave pulse sequences to tune the rotation angle $\varphi$, an arbitrary wave-particle superposition state can be prepared on demand. In the case of $\varphi=0$, the single phonon behaves completely as a particle, but as a wave for $\varphi=\pi/2$. The morphing between them can also be observed by tuning the rotation angle $\varphi$. The probability, $P_{k}$, of finding a phonon in the $k$th CNT is given by
\begin{equation}\label{eq:detection-probaility}
P_{k}=\frac{1}{2}+\left(-1\right)^{k}\frac{1}{2}\sin^{2}\left(\varphi\right)\cos\left(\phi\right),
\end{equation}
which includes two physical contributions, one from the particle
nature and the other from the wave nature. Note that the spin in a
mixed state
$\cos^{2}\left(\varphi\right)|0\rangle\langle0|+\sin^{2}\left(\varphi\right)|D\rangle\langle
D|$ is capable of reproducing the same measured statistics as in
Eq.~(\ref{eq:detection-probaility})~\cite{chaves2018causal}.
Thus, in order to exclude the classical interpretation and prove
the existence of the coherent wave-particle superposition, the
quantum coherence between the states $|0\rangle$ and $|D\rangle$
should be
verified~\cite{peruzzo2012quantum,kaiser2012entanglement,zheng2015quantum,liu2017twofold}.
Experimentally, such a verification can be implemented by
performing quantum state tomography to show all elements of the
density matrix of the spin~\cite{xing2017experimental}.

Next, we consider how to initialize and measure the mechanical system. Initially, the NV spin needs to be in the state $|D\rangle$ (i.e., the ground state of the spin qubit), one CNT, e.g., the first CNT, needs to be in its single-phonon state,  and the other CNT, e.g., the second CNT, needs to be in its vacuum state. To prepare such an initial state, we can begin with an arbitrary  state $\rho_{\rm
	ini}=\rho_{1}\otimes\rho_{2}\otimes\rho_{\rm spin}$, where
$\rho_{k}$ ($k=1,2$) and $\rho_{\rm spin}$ are the density matrices of the $k$th CNT
resonator and the spin, respectively.  One can apply a 532~nm laser pulse to initialize the spin
qubit in the state $|0\rangle$, and then apply a microwave
$\pi/2$-pulse to it, to obtain the superposition state
$\frac{1}{\sqrt{2}}\left(|0\rangle+|-1\rangle\right)$, which is
followed by a microwave $\pi$-pulse to obtain the spin qubit excited state
$|B\rangle$.
By using the sideband-cooling technique~\cite{xue2007cooling,you2008simultaneous,grajcar2008lower,ma2016cooling,clark2017sideband}, the CNT
resonators can be cooled down to their quantum ground state,
i.e., the acoustic vacuum $|\rm vac\rangle$. For example,
one can couple an auxiliary qubit with a large
spontaneous-emission rate to the CNT resonators~\cite{wang2017hybrid}. Once the
mechanical ground state is achieved, one can tune the spin-qubit
transition frequency $\omega_{q}$ to be close to the CNT resonance frequency $\omega_{m}$,
such that the spin-CNT coupling is then approximately given by a Jaynes-Cummings-type Hamiltonian
\begin{equation}
H_{\text{int}}\simeq\hbar g\left(\sigma_{+}b_{1}+\sigma_{-}b_{1}^{\dag}\right).
\end{equation}
When acting for a time equal to $\pi/\left(2g\right)$, such a Hamiltonian can, with the spin qubit in the excited state $|B\rangle$, transfer a mechanical excitation to the left CNT~\cite{o2010quantum}. Meanwhile, the spin qubit goes to its ground state $|D\rangle$. The desired initial state $|\Psi\rangle_{i}=\left(b_{1}^{\dag}\otimes\mathcal{I}_{2}|\rm vac\rangle\right)\otimes|D\rangle$ is then obtained.  For the phonon number measurement, we still need $\omega_{q}\simeq\omega_{m}$ as in the initialization, but the spin qubit is required to be in the ground state $|D\rangle$. In this situation, the Rabi frequency between the spin and the mechanical resonator depends on the number of phonons in the resonator~\cite{scully1997book,liu2004generation,hofheinz2008generation,hofheinz2009synthesizing,o2010quantum}. Thus by directly measuring the occupation probability of $|B\rangle$, the phonon number in each CNT can be obtained. The measurement of the spin state is enabled by the different fluorescence of the states $|0\rangle$ and $|\pm1\rangle$~\cite{doherty2013nitrogen}. To measure the state of the spin qubit, one can first apply a microwave $\pi$ pulse to map $|D\rangle\rightarrow\frac{1}{\sqrt{2}}\left(|0\rangle-|-1\rangle\right)$
and $|B\rangle\rightarrow\frac{1}{\sqrt{2}}\left(|0\rangle+|-1\rangle\right)$, and then apply a microwave $\pi/2$ pulse to map $\frac{1}{\sqrt{2}}\left(|0\rangle-|-1\rangle\right)\rightarrow|0\rangle$ and $\frac{1}{\sqrt{2}}\left(|0\rangle+|-1\rangle\right)\rightarrow|-1\rangle$. By measuring the Rabi oscillations between the states $|0\rangle$ and $|-1\rangle$ according to spin-state-dependent fluorescence~\cite{jelezko2004observation}, one can read out the spin qubit state. If one employs the repetitive-readout technique with auxiliary nuclear spins, the readout fidelity can be further improved~\cite{jiang2009repetitive}. 

\emph{Mechanical noise.}---Before discussing the mechanical noise, we need to analyze the total operation time, $\tau_{T}=2\tau_{0}+\tau_{1}$, required for our quantum delayed-choice experiment. Note that during $\tau_{T}$, we have neglected the spin single-qubit operation time due to the driving pulse length~$\sim$ ns~\cite{liu2015demonstration,liu2017single}. Since $0\leq\tau_{1}\leq2\pi/J$, we focus on the maximum $\tau_{T}$: $\tau_{T}^{{\rm max}}=5\pi/\left(2J\right)$. A modest spin-CNT coupling $g/2\pi=100$~kHz, which can be obtained by tuning the current $I$ and the distance $d$ (see Supplementary Section I), is able to mediate an effective CNT-CNT coupling $J/2\pi\simeq12$~kHz, thus giving $\tau_{T}^{{\rm max}}\simeq0.1$~ms. The relaxation time $T_{1}$ of a single NV spin at low temperatures can reach up to a few minutes. Moreover,  with spin echo techniques, a single spin in an ultra-pure diamond example typically has a dephasing time $T_{2}\simeq2$~ms even at room temperature~\cite{balasubramanian2009ultralong}, corresponding to a dephasing rate $\gamma_{s}/2\pi\simeq80$~Hz. When dynamical decoupling pulse sequences are employed, the dephasing time can be made even close to one second at low temperatures~\cite{bar2013solid}. These justify neglecting the spin decoherence. In this case, the mechanical noise dominates the dissipative processes. The dynamics of the system is therefore governed by the following master equation,
\begin{align}\label{eq:full-master-equation}
\dot{\rho}\left(t\right)=&\frac{i}{\hbar}\left[\rho\left(t\right),H\left(t\right)\right]-\frac{\gamma_{m}}{2}n_{\text{th}}
\sum_{k=1,2}\mathcal{L}\left(b_{k}^{\dag}\right)\rho\left(t\right)\nonumber\\
&-\frac{\gamma_{m}}{2}\left(n_{\text{th}}+1\right)\sum_{k=1,2}\mathcal{L}\left(b_{k}\right)\rho\left(t\right),
\end{align}
where $\rho\left(t\right)$ is the density operator of the system, $\gamma_{m}$ is the mechanical decay rate, $n_{\text{th}}=\left[\exp\left(\hbar\omega_{m}/k_{B}T\right)-1\right]^{-1}$ is the equilibrium phonon occupation at temperature $T$, and $\mathcal{L}\left(o\right)\rho\left(t\right)=o^{\dag}o\rho\left(t\right)-2o\rho\left(t\right)o^{\dag}+\rho\left(t\right)o^{\dag}o$ is the Lindblad superoperator. Here, $H\left(t\right)$ is a binary Hamiltonian of the form,
\begin{align}\label{eq:ternary-Hamiltonian}
H\left(t\right)=
    \begin{cases}
    H_{0}, & 0<t\leq\tau_{0}, \; {\rm and} \;\; \tau_{0}+\tau_{1}<t\leq\tau_{T}\\
    H_{1}, & \tau_{0}<t\leq\tau_{0}+\tau_{1},
    \end{cases}
\end{align}
with
\begin{equation}
H_{0}= J\left(\sum_{k=1,2}b_{k}^{\dag}b_{k}+b_{1}b_{2}^{\dag}+b_{2}b_{1}^{\dag}\right)\sigma_{z}
\end{equation}
and $H_{1}= Jb_{1}^{\dag}b_{1}\sigma_{z}$. In Eq. (\ref{eq:ternary-Hamiltonian}), we did not include the spin single-qubit operation before the third time interval because the length of the driving pulse is very short, as mentioned above. The master equation in Eq.~(\ref{eq:full-master-equation}) drives the phonon occupation of the $k$th CNT to be
\begin{align}
n_{k}&=\langle b_{k}^{\dag}b_{k}\rangle\left(\tau_{T}\right)\nonumber\\
&=P_{k}\exp\left(-\gamma_{m}\tau_{T}\right)+n_{\text{th}}\left[1-\exp\left(-\gamma_{m}\tau_{T}\right)\right],
\end{align}
at time $t=\tau_{T}$. For a realistic CNT, we can set the mechanical linewidth to be $\gamma_{m}/2\pi=0.4$~Hz~\cite{moser2014nanotube}, leading to a single-phonon lifetime of $\tau_{m}=1/\gamma_{m}\simeq400$~ms. In this situation, $\tau_{m}$ is much longer than the total operation time $\tau_{T}$, $\gamma_{m}\tau_{T}\ll1$ and, thus, we obtain
\begin{equation}\label{eq:finial_occupations_with_mechanical_noises}
n_{k}=P_{k}+n_{\text{th}}\gamma_{m}\tau_{T}.
\end{equation}
This shows that, in addition to the coherent signal $P_{k}$, the final occupation has a thermal contribution $n_{\text{th}}\gamma_{m}\tau_{T}$. In Fig. \ref{fig-morphing}, we demonstrate the morphing behavior between particle and wave at $T\simeq 10$~mK, according to Eq. (\ref{eq:finial_occupations_with_mechanical_noises}). To confirm this, we also plot numerical simulations, which are in exact agreement with our analytical expression. The thermal occupation, $n_{\text{th}}\gamma_{m}\tau_{T}$, increases as the phase $\phi$, because such a phase arises from the dynamical accumulation as discussed above. However, an extremely long phonon lifetime causes it to become negligible even at finite temperatures, as shown in Fig. \ref{fig-morphing}.

\begin{figure}[t]
\centering
\includegraphics[width=8.00cm]{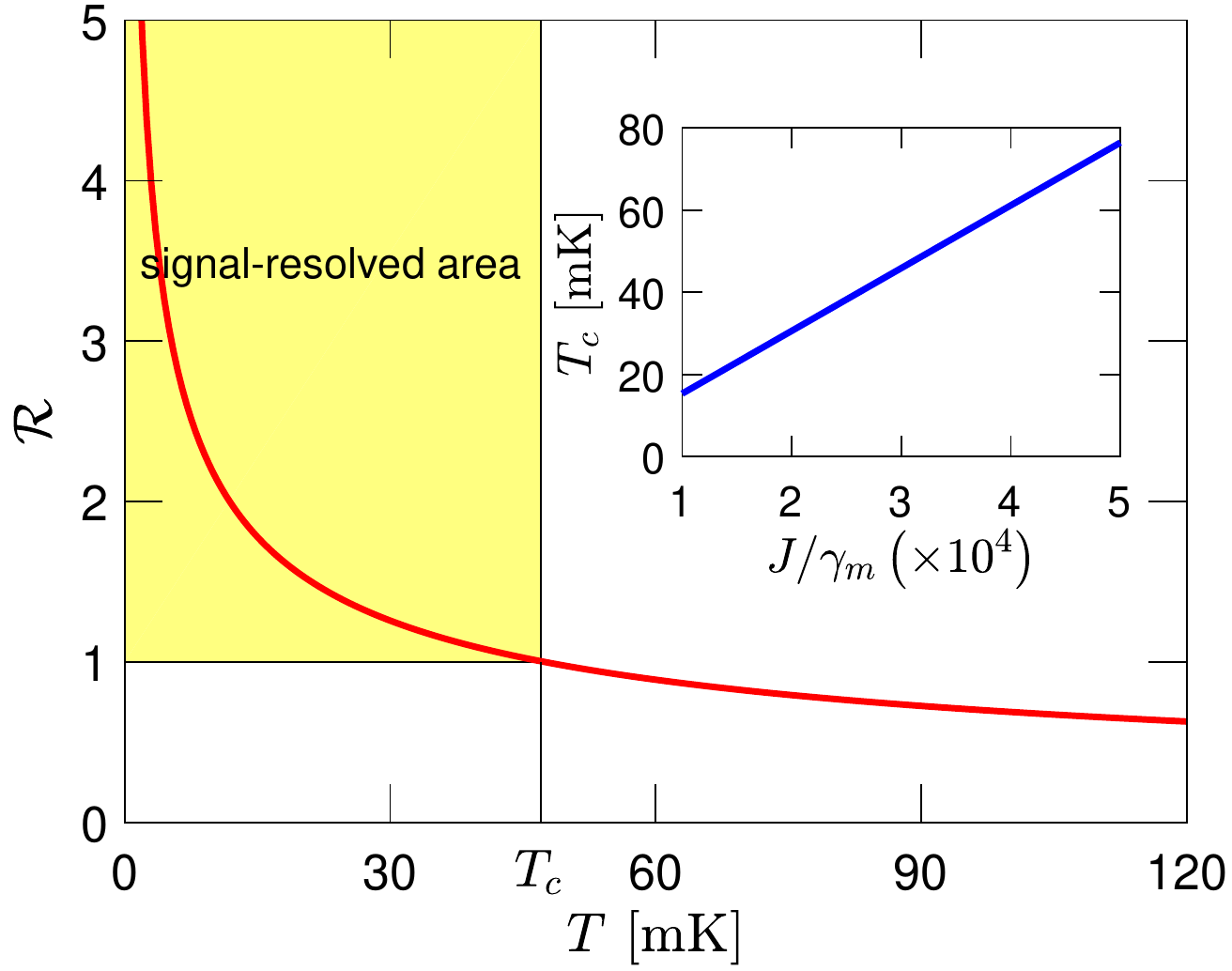}
\caption{Signal visibility $\mathcal{R}$ as a function of the temperature $T$. The yellow shaded area represents the signal-resolved regime, where the morphing between wave and particle can be effectively observed in the fluctuation noise. The vertical line corresponds to the critical temperature $T_{\rm c}$. The inset shows a linear increase in $T_{\rm c}$ with increasing the ratio of the spin-mediated CNT-CNT coupling strength $J$ to the mechanical-mode decay rate $\gamma_{m}$. Here, all parameters are set to be the same as in Fig.~\ref{fig-morphing}.}\label{fig-ratio-temp}
\end{figure}

We now consider the fluctuation noise. In the limit $\gamma_{m}\tau_{T}\ll1$, the fluctuation noise $\delta n_{k}^{{\rm noise}}$ in the phonon occupation $n_{k}$ is expressed, according to the analysis in the Supplementary Section IV, as
\begin{equation}
\left(\delta n_{k}^{{\rm noise}}\right)^{2}=P_{k}\left(2P_{k}-1\right)\gamma_{m}\tau_{m}+\left(2P_{k}+1\right)n_{{\rm th}}\gamma_{m}\tau_{T},
\end{equation}
where the first term is the vacuum fluctuation, which can be neglected, and the second term is the thermal fluctuation, which increases with temperature. To quantitatively describe the ability to resolve the coherent signal from the fluctuation noise, we typically employ the signal-to-noise ratio defined as
\begin{equation}
\mathcal{R}_{k}=\frac{P_{k}}{\delta n_{k}^{{\rm noise}}}.
\end{equation}
The signal-resolved regime often requires $\mathcal{R}_{k}>1$ for any $P_{k}$. However, the probability $P_{k}$ in the range zero to unity indicates that there always exist some $P_{k}$ such that $\mathcal{R}_{k}<1$, in particular, at finite temperatures. Nevertheless, we find that the total fluctuation noise
\begin{equation}
\mathcal{S}^{2}=\left(\delta n_{1}^{{\rm noise}}\right)^{2}+\left(\delta n_{2}^{{\rm noise}}\right)^{2}
\end{equation}
is kept below an upper bound
\begin{equation}
\mathcal{B}^{2}=\gamma_{m}\tau_{T}^{{\rm max}}+4n_{{\rm th}}\gamma_{m}\tau_{T}^{{\rm max}},
\end{equation}
and further that assuming $\mathcal{B}^{2}<1/2$ can make either or both of $\mathcal{R}_{1}$ and $\mathcal{R}_{2}$ greater than $1$. In this case, at least one CNT signal is resolved for each measurement. The conservation of the coherent phonon number equal to $1$ ensures that the unresolved signal can be inferred from the resolved one, which allows the morphing between wave and particle to be effectively observed from the fluctuation noise. To quantify this, we define a signal visibility as,
\begin{equation}
\mathcal{R}=\frac{\sqrt{2}}{2\mathcal{B}},
\end{equation}
in analogy to the signal-to-noise ratio $\mathcal{R}_{k}$. The ratio $\mathcal{R}$ describes the visibility of the total signal rather than the single CNT signals. At zero temperature ($n_{{\rm th}}=0$), the noise originates only from the vacuum fluctuation, and this yields $\mathcal{R}\gg1$. However, at finite temperatures, $n_{{\rm th}}$ increases as $T$, causing a decrease in $\mathcal{R}$, as shown in Fig.~\ref{fig-ratio-temp}. Therefore, the requirement of $\mathcal{R}>1$ sets an upper bound on the temperature, and as a result, leads to a critical temperature,
\begin{equation}\label{eq:Tc}
T_{\rm c}=\frac{\hbar\omega_{m}}{k_{B}\ln\left[\left(1+15\pi\gamma_{m}/J\right)/\left(1-5\pi\gamma_{m}/J\right)\right]}.
\end{equation}
The critical temperature linearly increases with $J/\gamma_{m}$, as plotted in the inset of Fig. \ref{fig-ratio-temp}. To increase $J$, we can increase the current $I$ through the CNTs, decrease the distance $d$ between the CNTs, or decrease the spin qubit transition frequency $\omega_{q}$. Furthermore, the increase in the CNT resonance frequency $\omega_{m}$ or the decrease in the CNT loss rate $\gamma_{m}$ can also lead to an increase in the critical temperature. For modest parameters of $J/2\pi=12$~kHz and $\gamma_{m}/2\pi=0.4$~Hz, a critical temperature $T_{\rm c}$ of $\simeq47$~mK, which is routinely accessible in current experiments, can be achieved.

\emph{Test of macroscopicity.---} We have described the
implementation of a quantum paradox with massive mechanical
objects with experimentally-distinguishable single-phonon
excitations. The question arises whether this proposal can be
considered as a test of macroscopicity~\cite{leggett2002testing,korsbakken2007measurement}. Typical proposals of such
tests (as cited below) have been based on implementing
superpositions of macroscopically distinguishable states of
classical-like systems, which are often referred to as
Schr\"odinger's cat states (see, e.g.,~\cite{Gerry05book}). Sometimes,
the meaning of Schr\"odinger's cat states is limited to
``superposition states of macroscopic systems, where the amplitude
of their excitations is large''~\cite{agarwal2013book}. Note, however,
that the term ``large amplitude'' can be understood in various
ways. These include the cases (criteria) when (i) the amplitudes
of the constituent states of a given superposition are large as in
classical systems or (ii) when these amplitudes are large enough
concerning their experimental distinguishability (i.e., compared
to the resolution of detectors). Strictly speaking, a state satisfying one of
these conditions, does \emph{not} necessarily satisfy the other.
For example, a superposition of coherent states, $\ket{\psi}
={\cal N}(\ket{\alpha}+\ket{\beta}$) with ${\cal N}$ being a
normalization constant, is a cat state according to criterion (i)
if $|\alpha|,|\beta|\gg 1$, but cannot be considered as a cat state
according to criterion (ii) if $\epsilon\equiv |\alpha-\beta| \ll
1$ is beyond the resolution of detectors. Conversely,
$\ket{\psi}$ is a cat state according to criterion (ii) if
$\epsilon$ can be resolved experimentally even if
$|\alpha|,|\beta|\approx 1$, i.e., when criterion (i) is not satisfied.
In the latter case, when the amplitude of such excitations is not
large in classical terms, but still macroscopically
distinguishable, the states are sometimes referred to as
Schr\"odinger's kitten states, as, e.g., those generated and
measured in Ref.~\cite{ourjoumtsev2006generating}. In this sense, the single-phonon
wave-particle superposition, given in Eq.~(\ref{eq:finial-state}), can be referred to
as a Schr\"odinger kitten state, since the excitations of the
macroscopic mechanical systems are small, i.e., at the
single-phonon level. Indeed, the amplitudes of single-phonon excitations are not	large enough to satisfy criterion (i). However, such
superpositions of single phonons are large enough that the
constituent states of the superposition, given in Eqs.~(\ref{eq:final-state-particle}) and~(\ref{eq:final-state-wave}), are experimentally distinguishable, thus satisfying criterion (ii). Therefore, such a test
of a quantum principle at the low-excitation level of massive
mechanical objects can also be viewed as a test at the macroscopic
scale, as claimed, e.g., in Refs.~\cite{hofer2016proposal,marinkovic2018optomechanical,ockeloen2018stabilized}
and references therein.

We note that a collective degree of freedom of many atoms does
	\emph{not} necessarily imply that the system is in a macroscopic
	quantum state. However, we showed that the studied system of
macroscopic resonators can be in a maximally entangled two-mode
state. This state is described by a non-positive Glauber-Sudarshan
$P$ function. This implies that the system itself is quantum.
Below we describe the method to amplify
the small-excitation kitten states, given in Eqs.~(\ref{eq:final-state-particle}) and~(\ref{eq:final-state-wave}),
to a cat state with large excitation.

\begin{figure}[t]
	\centering
	\includegraphics[width=8.2cm]{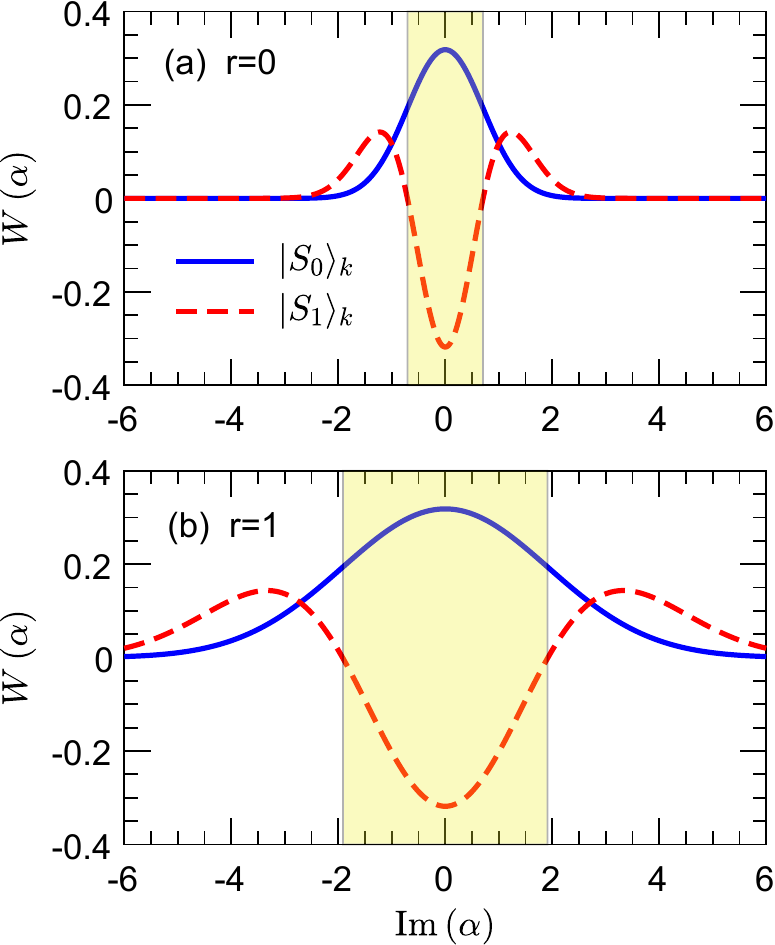}
	\caption{Wigner function for the states $|S_{0}\rangle_{k}=U_{k}|0\rangle_{k}$ (solid curves) and $|S_{1}\rangle_{k}=U_{k}|1\rangle_{k}$ (dashed curves) for $r=0$ in (a) and $1$ in (b). Here, $k=1,2$. The Wigner function is defined as $W\left(\alpha\right)=\pi^{-2}\int d^{2}\beta \exp\left(-i\beta\alpha^{*}-i\beta^{*}\alpha\right){\rm Tr}\left[\exp\left(i\beta^{*}a\right)\exp\left(i\beta a^{\dag}\right)\varrho\right]$. For both plots, we have assumed ${\rm Re}\left(\alpha\right)=0$. It is seen that the region marked in yellow, corresponding to the
		negative Wigner function for $|S_1\rangle_k$, increases
		with increasing squeezing parameter $r$.}\label{wigner_function}
\end{figure}

\emph{Amplification of the Schr\"odinger kitten states.---} Here
we apply the idea and method of Refs.~\cite{sekatski2009towards}
to show how to amplify the phonon numbers of the single-phonon
superposition states $|\text{particle}\rangle$ and
$|\text{wave}\rangle$, given in Eqs.~(\ref{eq:final-state-particle}) and~(\ref{eq:final-state-wave}), by squeezing
the mechanical modes $b_{1}$ and $b_{2}$. Thus, these states can
become Schr\"odinger's cat-like states. For simplicity, but
without loss of generality, here we consider a squeezing operator
\begin{align}
U_{k}=\exp\left[\tfrac {r}{2}
\left(b^{\dag2}_{k}-b_{k}^{2}\right)\right],
\end{align}
acting on the mode $b_{k}$ ($k=1,2$), with $r$ being a squeezing
parameter. This squeezing leads to 
\begin{align}
|S_{10}\rangle=\left(U_{1}b_{1}^{\dag}\otimes U_{2}\right)|{\rm vac}\rangle=&|S_{1}\rangle_{1}|S_{0}\rangle_{2},\\
|S_{01}\rangle=\left(U_{1}\otimes U_{2}b^{\dag}_{2}\right)|{\rm
	vac}\rangle =&|S_{0}\rangle_{1}|S_{1}\rangle_{2},
\end{align}
where we have defined the phonon squeezed Fock states
$|S_{0}\rangle_{k}=U_{k}|0\rangle_{k}$ and
$|S_{1}\rangle_{k}=U_{k}b^{\dag}_{k}|0\rangle_{k}$, with
$|0\rangle_{k}$ being the vacuum state of the mechanical mode
$b_{k}$. As a result, the states $|\text{particle}\rangle$ and
$|\text{wave}\rangle$  become
\begin{align}
|\mathcal{P}_{r}\rangle=&\frac{1}{\sqrt{2}}\left[\exp\left(i\phi\right)|S_{10}\rangle+i|S_{01}\rangle\right],\\
|\mathcal{W}_{r}\rangle=&\frac{1}{2}\left\{\left[\exp\left(i\phi\right)-1\right]|S_{10}\rangle
+i\left[\exp\left(i\phi\right)+1\right]|S_{01}\rangle\right\},
\end{align}
respectively. The final state $|\Psi\rangle_{f}$ becomes
\begin{equation}
|\Psi\rangle_{f}=\cos\left(\varphi\right)
|\mathcal{P}_{r}\rangle|0\rangle+\sin\left(\varphi\right)|\mathcal{W}_{r}\rangle|D\rangle.
\end{equation}
The modes $b_{k}$ for $k=1,2$ are transformed, via squeezing, to the
Bogoliubov modes described by
\begin{equation}
U_{k}^{\dag}b_{k}U_{k}=\cosh\left(r\right)b_{k}+\sinh\left(r\right)b_{k}^{\dag}.
\end{equation}
By using this unitary transformation, one obtains the average
phonon numbers of $|S_{0}\rangle_{k}$ and $|S_{1}\rangle_{k}$
equal to
\begin{align}
\leftidx{_{k}}\langle S_{0}|b^{\dag}_{k}b_{k}|S_{0}\rangle_{k}=&\sinh^{2}\left(r\right),\\
\leftidx{_{k}}\langle
S_{1}|b^{\dag}_{k}b_{k}|S_{1}\rangle_{k}=&3\sinh^{2}\left(r\right)+1.
\end{align}
We note that by applying this unconditional amplification method,
one can exponentially increase the distinguishability of the
states $|S_{10}\rangle$ and $|S_{01}\rangle$. Although, a
single-shot distinguishability of the mechanical-mode states
$|\mathcal{P}_{r}\rangle$ and $|\mathcal{W}_{r}\rangle$ is \emph{not}
increased, a tomographic distinguishability of these states in the
phase space is increased with the amplified amplitudes of the
mechanical-mode excitations. Indeed, the distinguishability of
$|\mathcal{P}\rangle_{r}$ and $|\mathcal{W}_{r}\rangle$, as measured by
the infidelity,
$\text{IF}=1-|\langle\mathcal{W}_{r}|\mathcal{P}_{r}\rangle|^2
=1-|\langle\text{wave}|\text{particle}\rangle|^2$, is independent
of the squeezing parameter $r$ for a given $\phi$. For any
$\phi\neq\pm\pi/2$, the states are distinguishable, and the
highest distinguishability is for $\phi=0,\pi$, for which the
infidelity is $\text{IF}=1/2$. Thus, even for such optimal values of
$\phi$, it is impossible to deterministically distinguish the
states $|\mathcal{P}_{r}\rangle$ and $|\mathcal{W}_{r}\rangle$ from each
other in a single-shot experiment. We refer to this property
as a single-shot distinguishability. Anyway, these mechanical
states can be macroscopically distinguished by performing, e.g., Wigner-function tomography on a number of their copies. Such
tomographic distinguishability in phase space indeed increases
with the squeezing parameter $r$, as shown in Fig.~\ref{wigner_function}.

Finally, we note that the famous optical prototypes of the
Schr\"odinger's cat states, which are given by the odd and even
coherent states, $\ket{\psi_{\pm}}={\cal N}( \ket{\alpha}\pm
\ket{-\alpha})$, \emph{cannot} be distinguished deterministically
in a single-shot experiment either. This is because the coherent
states $\ket{\alpha}$ and $\ket{-\alpha}$ are not orthogonal for
finite values of $\alpha$. Their overlap decreases exponentially
with increasing $\alpha$, so $\ket{\alpha}$ and $\ket{-\alpha}$
become orthogonal in the limit of large $|\alpha|$. However, this
amplification of $\alpha$ cannot be done deterministically,
because this process is prohibited by the no-cloning and
no-signalling theorems. Indeed, non-orthogonal states cannot be
deterministically transformed to orthogonal (thus, completely
distinguishable) states. Note that popular methods of amplifying
small-amplitude states are based on either (i) probabilistic but
accurate amplification or (ii) deterministic but inaccurate
cloning. For example, the method described, e.g., in Refs.~\cite{ourjoumtsev2006generating,lund2004conditional}
is probabilistic, because it is based on
conditional measurements performed on two copies of
$\ket{\psi_{\pm}}$. In contrast to this, the amplification method
in Ref.~\cite{sekatski2009towards}, as applied here, corresponds to approximate
quantum cloning, i.e., phase-covariant cloning by stimulated
emission.

\section{Discussion}
We have presented a proposal for a quantum delayed-choice experiment with nanomechanical resonators, which enables a macroscopic test of an arbitrary quantum wave-particle superposition. The ability to tolerate the mechanical noise has also been given here, demonstrating that our proposal can be implemented with current experimental techniques. While we have chosen to focus on a spin-nanomechanical setup, the present method could be directly extended to other hybrid systems, for example, mechanical devices coupled to a superconducting atom~\cite{xiang2013hybrid,o2010quantum,gu2017microwave}. 
Recently, an experimental work reported that photons can be
entangled in their wave-particle degree of freedom~\cite{rab2017entanglement}. This
indicates that the wave-particle nature of photons may be used to
encode flying qubits for long-distance quantum communication.
Photons are ideal quantum information carriers, but they are
difficult to store. In contrast to photons, long-lived phonons
could be used for optical information storage~\cite{fiore2011storing}. Our study
shows that phonons can also be prepared in a wave-particle
superposition state, and that the wave-particle nature of phonons
is not more special than their other degrees of freedom. Thus, the
wave-particle degree of freedom of phonons may be exploited for
storing quantum information encoded in the wave-particle degree of
freedom of photons. In addition, optomechanical interactions can
couple a mechanical mode to optical modes at different
frequencies~\cite{dong2012optomechanical}. Thus, the mechanical wave-particle degree of
freedom may be employed to map quantum information encoded in the
wave-particle degree of freedom from photons at a given frequency
to photons at any desired frequency. The
mechanical wave-particle nature, as a new degree of freedom, may
find various applications in quantum information.

We believe that the macroscopicity of our
	single-phonon wave-particle superposition is highly
	counter-intuitive, as based on a refined version of the quantum
	paradox, even if the mechanical resonators are in the
	single-phonon-excitation regime. Indeed, we analyzed a ``nested''
	kitten state, as given in Eq.~(\ref{eq:finial-state}), where the particle and wave
	states, given in Eqs.~(\ref{eq:final-state-particle}) and~(\ref{eq:final-state-wave}), are purely mechanical kitten
	states for $\phi\neq \pm \pi/2$. Moreover, we have described a method, based on mechanical-mode
	squeezing, which enables the amplification of small-excitation
	Schr\"odinger kitten states, given in Eqs.~(\ref{eq:final-state-particle}) and~(\ref{eq:final-state-wave}), to
	large-excitation Schr\"odinger cat states of the massive
	mechanical resonators. For these reasons, an experimental realization of our
	proposal can be a fundamental test of a coherent wave-particle
	superposition of massive objects with phonon excitations, which
	can be increased exponentially by squeezing.  Hence, this proposed quantum delayed-choice experiment of massive mechanical resonators not only leads to a better understanding of quantum theory at the macroscopic scale, but also indicates that, like the vertical and horizontal polarizations of photons, the mechanical wave-particle nature, as an additional degree of freedom of phonons, may be widely exploited for quantum information applications.

\section{data availability}
The data that supports the findings of this study are
available in the Supplementary Information file. Additional data are also available from the corresponding
authors upon reasonable request.

\section{Acknowledgments}
W.Q. thanks Peng-Bo Li for valuable discussions.
W.Q. and J.Q.Y. were supported in part by the National Key Research and
Development Program of China (Grant No. 2016YFA0301200),
the China Postdoctoral Science Foundation (Grant No. 2017M610752),
and the NSFC (Grant No. 11774022). G.L.L. is supported in part by National Key Research and Development Program of China (2017YFA0303700);
Beijing Advanced Innovation Center for Future Chip (ICFC). A.M. and
F.N. acknowledge the support of a grant from the John Templeton
Foundation. F.N. is supported in part by the:
MURI Center for Dynamic Magneto-Optics via the
Air Force Office of Scientific Research (AFOSR) (FA9550-14-1-0040),
Army Research Office (ARO) (Grant No. Grant No. W911NF-18-1-0358),
Asian Office of Aerospace Research and Development (AOARD) (Grant No. FA2386-18-1-4045),
Japan Science and Technology Agency (JST) (Q-LEAP program, ImPACT program, and CREST Grant No. JPMJCR1676),
Japan Society for the Promotion of Science (JSPS) (JSPS-RFBR Grant No. 17-52-50023, and JSPS-FWO Grant No. VS.059.18N),
and RIKEN-AIST Challenge Research Fund.

\section{Author contributions}
W.Q. and A.M. developed the theory and performed
the calculations. G.L.L., J.Q.Y., and F. N. supervised the project. All authors researched, collated, and wrote this paper.

\section{additional information}
{\bf Supplementary Information} accompanies the paper on the npj
Quantum Information website (doi:xxx).

{\bf Competing interests:} The authors declare no competing interests.

\bibliographystyle{naturemag}

\clearpage \widetext
\begin{center}
	\section{\large Supplementary information}
\end{center}
\setcounter{equation}{0} \setcounter{figure}{0}
\setcounter{table}{0} \setcounter{page}{1} \setcounter{secnumdepth}{3}\makeatletter
\renewcommand{\theequation}{S\arabic{equation}}
\renewcommand{\thefigure}{S\arabic{figure}}
\renewcommand{\bibnumfmt}[1]{[S#1]}
\renewcommand{\citenumfont}[1]{S#1}
\renewcommand\thesection{S\arabic{section}}

\def\@hangfrom@section#1#2#3{\@hangfrom{#1#2#3}}
\makeatother

\begin{quote}
	Here, we, first, in Sec.~\ref{sec:Spin-controlled coherent coupling between separated mechanical resonators} present more details of how to obtain the spin-controlled coherent coupling between separated mechanical resonators. Second, in Sec.~\ref{sec:controlled Hadamard gate}, we show the detailed implementation of the controlled Hadamard gate, the phase gate, and the mechanical quantum delayed-choice experiment.
	Next, in Sec.~\ref{sec:Phonon occupations at finite temperature}, we derive in detail the phonon occupation of each CNT at finite temperatures. Then, Sec.~\ref{sec:Signal-to-noise ratio at finite temperature} describes the detailed derivation of the fluctuation noise and the detailed analysis of the requirement of resolving the coherent signal from the environment-induced fluctuation. Finally, in Sec.~\ref{sec:numerical simulations} we show the method of the numerical simulation used in this work.
\end{quote}

\section{Spin-controlled coherent coupling between separated mechanical resonators}
\label{sec:Spin-controlled coherent coupling between separated mechanical resonators}
\begin{figure}[tbph]
	\centering
	\includegraphics[width=14.0cm]{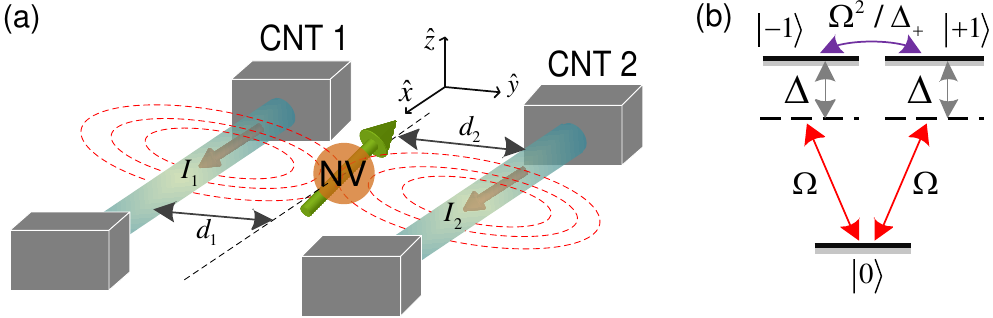}
	\caption{(Color online) (a) Schematic representation of a mechanical quantum delayed-choice experiment with an NV electronic spin and two carbon nanotubes (CNTs). The mechanical vibrations of the CNTs, labelled by $k=1,2$, are completely decoupled or coherently coupled, depending, respectively, on whether or not the intermediate spin is in the spin state $|0\rangle$, with the dc current $I_{k}$ through the $k$th CNT, and the distance $d_{k}$ between the spin and the $k$th CNT. (b) Level structure of the driven NV spin in the electronic ground state. Here we have assumed that the Zeeman splitting between the spin states $|\pm1\rangle$ is eliminated by applying an external field. }\label{sfig:schematic}
\end{figure}
The effective Hamiltonian $H_{\text{eff}}$ in the article describes a spin-mediated CNT-CNT coupling conditioned on the NV spin state. This is the basic element underlying our proposal. To understand more explicitly the spin-controlled coupling between the CNTs, in this section we derive in detail the effective Hamiltonian. We consider a hybrid quantum system consisting of two parallel CNTs and an NV electronic spin (a qutrit), as depicted in Fig.~\ref{sfig:schematic}(a). Here, for convenience, illustrations in Figs. 1(b) and 1(c) in the article are reproduced in Figs.~\ref{sfig:schematic}(b) and ~\ref{sfig:schematic}(a), respectively. The CNTs, respectively, carry dc currents $I_{1}$ and $I_{2}$, both along the $+\hat{x}$--direction. A spin is placed between them, at a distance $d_{1}$ ($d_{2}$) from the first (second) CNT. According to the Biot-Savart law, the CNTs can, at the position of the spin, generate a magnetic field $\vec{B}_{\text{cnt}}^{\left(0\right)}=B_{\text{cnt}}^{\left(0\right)}\hat{z}$, where
\begin{equation}\label{seq:zero-order-field}
B_{\text{cnt}}^{\left(0\right)}=\sum_{k=1,2}\left(-1\right)^{k-1}\frac{\mu_{0}I_{k}}{2\pi d_{k}},
\end{equation}
$\hat{\epsilon}$ ($\epsilon=x,y,z$) is a unit vector in the $\hat{\epsilon}$--direction, $\mu_{0}$ is the vacuum permeability, and the subscript ``cnt'' refers to the CNTs. When the CNTs vibrate along the $\hat{y}$-direction, the magnetic field is parametrically modulated by their mechanical displacements $y_{1}$ and $y_{2}$, and then is reexpressed, up to first order, as $\vec{B}_{\text{cnt}}=\vec{B}_{\text{cnt}}^{\left(0\right)}+\vec{B}_{\text{cnt}}^{\left(1\right)}$, where $\vec{B}_{\text{cnt}}^{\left(1\right)}=B_{\text{cnt}}^{\left(1\right)}\hat{z}$ is a first-order modification, and where $B_{\text{cnt}}^{\left(1\right)}=\sum_{k=1,2}G_{k}y_{k}$, with a magnetic-field gradient,
\begin{equation}\label{seq:field-shift-per-displacement}
G_{k}=\frac{\mu_{0}I_{k}}{2\pi d_{k}^{2}}.
\end{equation}
Note that, here, $y_{1}>0$ ($y_{2}<0$) indicates a decrease in $d_{1}$ ($d_{2}$). Therefore, the sign, $\left(-1\right)^{k-1}$, in Eq.~(\ref{seq:zero-order-field}) does not appear in Eq.~(\ref{seq:field-shift-per-displacement}). Furthermore, an external magnetic field, $\vec{B}_{\text{ext}}=B_{x}\left(t\right)\hat{x}+B_{z}\hat{z}$, is applied to the NV spin. We have assumed, as required below, that $B_{x}\left(t\right)$ is a time-dependent component but $B_{z}$ is a dc component. The Hamiltonian governing the NV spin is therefore given by
\begin{equation}
H_{\text{NV}}=\hbar DS_{z}^{2}+\mu_{B}g_{s}\left[B^{\left(0\right)}_{\text{cnt}}+B_{z}\right]S_{z}
+\mu_{B}g_{s}B_{x}\left(t\right)S_{x}+\mu_{B}g_{s}B_{\text{cnt}}^{\left(1\right)}S_{z},
\end{equation}
where $g_{s}\simeq2$ is the Land\'{e} factor, $\mu_{B}$ the Bohr magneton, $D\simeq2\pi\times2.87$~GHz the zero-field splitting, and $S_{\epsilon}$ the $\epsilon$--component of the spin operator $\vec{S}$ ($\epsilon=x,y,z$). In terms of the eigenstates, $\left\{|m_{s}\rangle,m_{s}=0,\pm1\right\}$, of $S_{z}$, the operator $S_{\epsilon}$ is expanded as
\begin{align}
S_{x}=\frac{1}{2}\left(
\begin{array}{ccc}
0 & \sqrt{2} & 0 \\
\sqrt{2}  &  0 & \sqrt{2} \\
0 & \sqrt{2} & 0 \\
\end{array}
\right), \;
S_{y}=\frac{1}{2i}\left(
\begin{array}{ccc}
0 & \sqrt{2} & 0 \\
-\sqrt{2}  &  0 & \sqrt{2} \\
0 & -\sqrt{2} & 0 \\
\end{array}
\right), \; {\rm and} \;\;
S_{z}=\frac{1}{2}\left(
\begin{array}{ccc}
+1 & 0 & 0 \\
0  &  0 & 0 \\
0 & 0 & -1 \\
\end{array}
\right),
\end{align}
and accordingly, the Hamiltonian $H_{\rm NV}$ is transformed to
\begin{align}\label{seq:NV-Hamiltonian-expressed-in-Sz-basis}
H_{\text{NV}}=&\left\{\hbar D+\mu_{B}g_{s}\left[B^{\left(0\right)}_{\text{cnt}}
+B_{z}\right]\right\}|+1\rangle\langle+1|+\left\{\hbar D-\mu_{B}g_{s}\left[B^{\left(0\right)}_{\text{cnt}}
+B_{z}\right]\right\}|-1\rangle\langle-1|\nonumber\\
&+\frac{1}{\sqrt{2}}\mu_{B}g_{s}B_{x}\left(t\right)\left(|-1\rangle\langle 0|+|+1\rangle\langle 0|+\text{H.c.}\right)\nonumber\\
&+\mu_{B}g_{s}B_{\text{cnt}}^{\left(1\right)}\left(|+1\rangle\langle+1|-|-1\rangle\langle-1|\right).
\end{align}
We find that the magnetic field along the $\hat{z}$--direction causes different Zeeman shifts to be imposed, respectively, on the spin states $|\pm1\rangle$, and also that the magnetic field along the $\hat{x}$--direction drives the transition between the spin states $|0\rangle$ and$|\pm1\rangle$.

The quantum treatment of the mechanical motion demonstrates that the mechanical vibrations of the CNTs can be modelled by two single-mode harmonic oscillators with a Hamiltonian
\begin{equation}\label{seq:quantized-nanotube-selfHamiltonian}
H_{\text{mv}}=\sum_{k=1,2}\hbar\omega_{k}b_{k}^{\dag}b_{k},
\end{equation}
where $\omega_{k}$ is the phonon frequency and $b_{k}$ ($b_{k}^{\dag}$) is the phonon annihilation (creation) operator. Here, we have subtracted the constant zero-point energy $\hbar\omega_{k}/2$. The mechanical displacement $y_{k}$ is accordingly expressed as
\begin{equation}\label{seq:quantized-mechanical-amplitude}
y_{k}=y_{\text{zp}}^{\left(k\right)}\left(b_{k}+b_{k}^{\dag}\right)
\equiv y_{\text{zp}}^{\left(k\right)}q_{k},
\end{equation}
where $q_{k}$ is the canonical phonon position operator, and $y_{\text{zp}}^{\left(k\right)}=\left[\hbar/\left(2m_{k}\omega_{k}\right)\right]^{1/2}$, with $m_{k}$ being the effective mass, describes the zero-point (zp) motion. Combining Eqs.~(\ref{seq:NV-Hamiltonian-expressed-in-Sz-basis}), (\ref{seq:quantized-nanotube-selfHamiltonian}), and (\ref{seq:quantized-mechanical-amplitude}) gives the full Hamiltonian of the hybrid system,
\begin{align}\label{seq:original-full-Hamiltonian}
H_{F}=&\sum_{k=1,2}\hbar\omega_{k}b_{k}^{\dag}b_{k}+\left\{\hbar D+\mu_{B}g_{s}\left[B^{\left(0\right)}_{\text{cnt}}
+B_{z}\right]\right\}|+1\rangle\langle+1|\nonumber\\
&+\left\{\hbar D-\mu_{B}g_{s}\left[B^{\left(0\right)}_{\text{cnt}}
+B_{z}\right]\right\}|-1\rangle\langle-1|\nonumber\\
&+\frac{1}{\sqrt{2}}\mu_{B}g_{s}B_{x}\left(t\right)\left(|-1\rangle\langle 0|+|+1\rangle\langle 0|+\text{H.c.}\right)\nonumber\\
&+\sum_{k=1,2}\mu_{B}g_{s}G_{k}y_{\text{zp}}^{\left(k\right)}\left(|+1\rangle\langle+1|-|-1\rangle\langle-1|\right)q_{k}.
\end{align}
The last line in Eq.~(\ref{seq:original-full-Hamiltonian}) describes a magnetic coupling between the spin and the mechanical modes. In order to realize a tunable detuning between them, $B_{x}\left(t\right)$ is chosen to be $B_{x}\left(t\right)=B_{0}\cos\left(\omega_{0}t\right)$ with amplitude $B_{0}$ and frequency $\omega_{0}$. In a frame rotating at $H_{\text{rot}}=\hbar\omega_{0}\left(|-1\rangle\langle-1|+|+1\rangle\langle+1|\right)$, the full Hamiltonian can be divided into two parts, $H_{F}=H_{\text{low}}+H_{\text{high}}$, where
\begin{align}
H_{\text{low}}=&\sum_{k=1,2}\hbar\omega_{k}b_{k}^{\dag}b_{k}+\hbar\delta_{+}|+1\rangle\langle+1|+\hbar\delta_{-}|-1\rangle\langle-1|\nonumber\\
&+\hbar\Omega\left(|-1\rangle\langle 0|+|+1\rangle\langle 0|+\text{H.c.}\right)\nonumber\\
&+\sum_{k=1,2}\hbar g_{k}\left(|+1\rangle\langle+1|-|-1\rangle\langle-1|\right)q_{k},\\
H_{\text{high}}=&\hbar\Omega\left[\exp\left(i2\omega_{0}t\right)|-1\rangle\langle 0|+\exp\left(i2\omega_{0}t\right)|+1\rangle\langle 0|+\text{H.c.}\right],
\end{align}
account for the low- and high-frequency components, respectively. Here, we have defined
\begin{align}
\hbar\delta_{\pm}&=\hbar D\pm\mu_{B}g_{s}\left[B^{\left(0\right)}_{\text{nt}}+B_{z}\right]-\hbar\omega_{0},\nonumber\\
\hbar\Omega&=\frac{1}{2\sqrt{2}}\mu_{B}g_{s}B_{0},\nonumber\\
\hbar g_{k}&=\mu_{B}g_{s}G_{k}y_{\text{zp}}^{\left(k\right)}.
\end{align}
Roughly, having $\delta_{\pm}^{\prime}=\delta_{\pm}+2\omega_{0}\gg\Omega$ allows one to make the rotating-wave approximation (RWA), and to straightforwardly remove $H_{\text{high}}$. However, as demonstrated in Sec.~\ref{sec:numerical simulations}, the accumulated error increases during the evolution, causing the dynamics driven by $H_{\text{low}}$ to deviate largely from that driven by $H_{F}$. Thus, we are \emph{not} using the RWA here. In order to suppress the error accumulation, we need to analyze the effects of $H_{\text{high}}$ in the limit $\delta_{\pm}^{\prime}\gg\Omega$. In such a limit, we can employ a time-averaging treatment for the high-frequency component $H_{\text{high}}$~\cite{gamel2010time,qin2018exponentially}, and as a result, its effective behavior is described by the following time-averaged Hamiltonian,
\begin{align}
\overline{H}_{\text{high}}=&\hbar\left(\frac{2\Omega^{2}}{\delta^{\prime}_{-}}
+\frac{\Omega^{2}}{\delta^{\prime}_{+}}\right)|-1\rangle\langle-1|
+\hbar\left(\frac{\Omega^{2}}{\delta^{\prime}_{-}}+\frac{2\Omega^{2}}{\delta^{\prime}_{+}}
\right)|+1\rangle\langle+1|\nonumber\\
&+\hbar\frac{\Omega^{2}}{2}\left(\frac{1}{\delta_{-}^{\prime}}+\frac{1}{\delta_{+}^{\prime}}\right)
\left\{\exp\left[i\left(\delta_{-}+\delta_{-}^{\prime}-\delta_{+}-\delta_{+}^{\prime}\right)t\right]|-1\rangle\langle+1|+\text{H.c.}\right\},
\end{align}
where the first line corresponds to the energy shifts of the spin states $|\pm1\rangle$, and the second line describes a coherent coupling between these. Accordingly, the full Hamiltonian $H_{F}$ is approximated to be a time-independent form,
\begin{equation}\label{seq:time-independent-full-Hamiltonian}
H_{F}\simeq H_{\text{low}}+\overline{H}_{\text{high}}.
\end{equation}
As seen in Sec.~\ref{sec:numerical simulations}, the error accumulation is strongly suppressed when $\overline{H}_{\text{high}}$ is included.

Tuning $B^{\left(0\right)}_{\text{cnt}}+B_{z}=0$ yields $\delta_{+}=\delta_{-}=\Delta_{-}$ and $\delta_{+}^{\prime}=\delta_{-}^{\prime}=\Delta_{+}$, implying that the spin states $|\pm1\rangle$ have the same Zeeman shift of $\Delta=\Delta_{-}+3\Omega^{2}/\Delta_{+}$, as shown in Fig. \ref{sfig:schematic}(b). Therefore, we can define a bright state, $|B\rangle=\left(|+1\rangle+|-1\rangle\right)/\sqrt{2}$, which is dressed by the spin state $|0\rangle$, and a dark state,
$|D\rangle=\left(|+1\rangle-|-1\rangle\right)\sqrt{2}$, which decouples from the spin state $|0\rangle$.
In terms of the states $|B\rangle$ and $|D\rangle$, the full Hamiltonian becomes
\begin{align}\label{seq:full-Hamiltonian-in-terms-of-bright-and-dark-states}
H_{F}\simeq&\sum_{k=1,2}\hbar\omega_{k}b_{k}^{\dag}b_{k}+\hbar\Delta\left(|B\rangle\langle B|+|D\rangle\langle D|\right)+\hbar\sqrt{2}\Omega\left(|0\rangle\langle B|+|B\rangle\langle 0|\right)\nonumber\\
&+\sum_{k=1,2}\hbar g_{k}\left(|B\rangle\langle D|+|D\rangle\langle B|\right)q_{k}+\hbar\frac{\Omega^{2}}{\Delta_{+}}\left(|B\rangle\langle B|-|D\rangle\langle D|\right).
\end{align}
The dressing mechanism allows us to introduce two dressed states,
\begin{align}
|\Phi_{-}\rangle=&\cos\left(\theta\right)|0\rangle-\sin\left(\theta\right)|B\rangle,\\ |\Phi_{+}\rangle=&\sin\left(\theta\right)|0\rangle+\cos\left(\theta\right)|B\rangle,
\end{align}
where $\tan\left(2\theta\right)=2\sqrt{2}\Omega/\Delta$. Upon substituting them back into the full Hamiltonian in Eq.~(\ref{seq:full-Hamiltonian-in-terms-of-bright-and-dark-states}) and then using the identity operator
$\mathcal{I}=|D\rangle\langle D|+|\Phi_{-}\rangle\langle\Phi_{-}|+|\Phi_{+}\rangle\langle\Phi_{+}|$, we can straightforwardly obtain
\begin{align}
H_{F}\simeq&\sum_{k=1,2}\hbar\omega_{k}b_{k}^{\dag}b_{k}+\hbar\omega_{+}|\Phi_{+}\rangle\langle\Phi_{+}|+\hbar\omega_{D}|D\rangle\langle D|\nonumber\\
&+\sum_{k=1,2}\hbar\left[g_{k}^{\left(-\right)}|\Phi_{-}\rangle\langle D|+g_{k}^{\left(+\right)}|D\rangle\langle\Phi_{+}|+\text{H.c.}\right]q_{k}\nonumber\\
&+\hbar\frac{\Omega^{2}}{\Delta_{+}}\left[\cos\left(2\theta\right)|\Phi_{+}\rangle\langle\Phi_{+}|
-\frac{1}{2}\sin\left(2\theta\right)\left(|\Phi_{+}\rangle\langle\Phi_{-}|+\text{H.c.}\right)
-\cos^{2}\left(\theta\right)|D\rangle\langle D|\right].
\end{align}
Here,
\begin{align}
\omega_{+}=&\sqrt{\Delta^{2}+8\Omega^{2}},\\
\omega_{D}=&\frac{1}{2}\left(\Delta+\sqrt{\Delta^{2}+8\Omega^{2}}\right),\\
g_{k}^{\left(-\right)}=&-g_{k}\sin\left(\theta\right),\\
g_{k}^{\left(+\right)}=&g_{k}\cos\left(\theta\right).
\end{align}
Under the assumption of $\Delta\gg\Omega$, we have $\theta\simeq0$, such that $\sin\left(\theta\right)\simeq\sin\left(2\theta\right)\simeq0$, $\cos\left(\theta\right)\simeq\cos^{2}\left(\theta\right)\simeq\cos\left(2\theta\right)\simeq1$, $\omega_{+}\simeq\Delta+4\Omega^2/\Delta$, $\omega_{D}\simeq\Delta+2\Omega^{2}/\Delta$, and $|\Phi_{+}\rangle\simeq|B\rangle$. In this limit, the coupling between $|0\rangle$ and $|B\rangle$ only causes an energy splitting, of $\simeq2\Omega^{2}/\Delta$, between the states $|B\rangle$ and $|D\rangle$, so $|B\rangle$ and $|D\rangle$ can be used to define a spin qubit. Correspondingly, the full Hamiltonian is approximated as
\begin{align}\label{seq:effective-full-Hamiltonian}
H_{F}^{\prime}=\sum_{k=1,2}\hbar\omega_{k}b_{k}^{\dag}b_{k}+\frac{1}{2}\hbar\omega_{q}\sigma_{z}+\sum_{k=1,2}\hbar g_{k}\sigma_{x}q_{k},
\end{align}
where $\omega_{q}=2\Omega^{2}/\Delta+2\Omega^{2}/\Delta_{+}$, $\sigma_{z}=|B\rangle\langle B|-|D\rangle\langle D|$, and $\sigma_{x}=\sigma_{+}+\sigma_{-}$ with $\sigma_{-}=|D\rangle\langle B|$ and $\sigma_{+}=\sigma_{-}^{\dag}$. Modest parameters~\cite{sazonova2004tunable,ustunel2005modeling,garcia2007mechanical,ning2014transversally,truax2014axially,tsioutsios2017real,liu2019sensing}, $m_{k}=1.0\times10^{-22}$~kg, $\omega_{k}/2\pi=2$~MHz, $d_{k}\simeq2$~nm, and $I_{k}\simeq380$~$\text{nA}$, could result in a spin-CNT coupling of up to $g_{k}/2\pi\simeq100$~kHz.

Furthermore, from Eq. (\ref{seq:effective-full-Hamiltonian}) it is found that the sequential actions of the terms $\sigma_{+}b_{1}$ and $\sigma_{-}b_{2}^{\dag}$, as well as of the counter-rotating terms $\sigma_{-}b_{1}$ and $\sigma_{+}b_{2}^{\dag}$, can transfer a mechanical phonon from the left to the right CNT, and the reverse process is caused by their Hermitian conjugates. When restricting our discussion to a dispersive regime,
\begin{equation}\label{seq:limit-for-disperse-regime}
\omega_{q}\pm\omega_{k}\gg |g_{k}|,
\end{equation}
this phonon transfer becomes dominant. Hence, in the dispersive regime the dynamics described by $H^{\prime}_{F}$ in Eq.~(\ref{seq:effective-full-Hamiltonian}) enables a spin quantum bus for the mechanical phonons and can be used to realize a coherent CNT-CNT coupling. In order to show more explicitly, we rewrite $H_{F}^{\prime}$ in the interaction picture as
\begin{equation}\label{seq:effective-full-Hamiltonian-in-interaction-picture}
H_{F}^{\prime}=\sum_{k=1,2}\hbar g_{k}\left\{\sigma_{+}b_{k}\exp\left[i\left(\omega_{q}-\omega_{k}\right)\right]
+\sigma_{+}b_{k}^{\dag}\exp\left[i\left(\omega_{q}+\omega_{k}\right)\right]+\text{H.c.}\right\}.
\end{equation}
The condition in Eq.~(\ref{seq:limit-for-disperse-regime}) justifies to use a time-averaging treatment of the Hamiltonian $H_{F}^{\prime}$~\cite{gamel2010time,qin2018exponentially}. In the time-averaging treatment, all terms in Eq.~(\ref{seq:effective-full-Hamiltonian-in-interaction-picture}) are considered as high-frequency components and exhibit time-averaged behaviors. Based on this, the dynamics of the system can be determined by an effective Hamiltonian
\begin{align}\label{seq:effective-Hamiltonian}
H_{\text{eff}}=\frac{2\hbar\omega_{q}}{\omega_{q}^{2}-\omega_{m}^{2}}
\left[\sum_{k=1,2}g_{k}^{2}b_{k}^{\dag}b_{k}+g_{1}g_{2}\left(b_{1}b_{2}^{\dag}+b_{2}b_{1}^{\dag}\right)\right]\otimes \sigma_{z}.
\end{align}
Here, we have assumed that $\omega_{k}=\omega_{m}$. As expected, Eq.~(\ref{seq:effective-Hamiltonian}) shows a coherent spin-mediated CNT-CNT coupling, corresponding to the standard linear coupler transformation, which can give rise to a direct phonon exchange. Thus in this case, the spin qubit works as a quantum bus. At the same time, it also shows that the CNT-CNT coupling can be turned off if the intermediate spin is in the state $|0\rangle$. This is because the NV spin in the state $|0\rangle$ is decoupled from the CNTs, and the mechanical phonons can no longer be transferred from one CNT to another. Specifically, if the spin is in the state $|D\rangle$ or $|B\rangle$, the CNTs are coupled; however, if the spin is instead in the state $|0\rangle$, they are decoupled. Note that in Eq.~(\ref{seq:effective-Hamiltonian}) ac Stark shifts caused to be imposed on the qubit have been excluded because we focus only on the quantum states of the CNTs.

In the last part of this section, we evaluate the direct coupling between the CNTs. For simplicity, we assume that $I_{k}=I$, $d_{k}=d$, and that the CNTs have the same length $L$. The attractive force acting on the $k$th CNT is
\begin{equation}
\vec{F}_{k}=\left(-1\right)^{k-1}F\hat{y},
\end{equation}
where
\begin{equation}
F=\frac{\mu_{0}LI^{2}}{2\pi\left(d-y_{1}+y_{2}\right)}
\end{equation}
is the force size. The work done by the force is given straightforwardly by
\begin{align}
W=\frac{\mu_{0}LI^{2}\left(y_{1}-y_{2}\right)}{2\pi\left(d-y_{1}-y_{2}\right)}.
\end{align}
After applying a perturbation expansion and then a quantization, this direct CNT-CNT coupling is found to be
\begin{align}
W=&\hbar W^{\left(1\right)}\left(b_{1}-b_{2}+\text{H.c.}\right)\nonumber\\
&+\hbar W^{\left(2\right)}\left[\left(b_{1}+b_{1}^{\dag}\right)^{2}+\left(b_{2}+b_{2}^{\dag}\right)^{2}
-2\left(b_{1}+b_{1}^{\dag}\right)\left(b_{2}+b_{2}^{\dag}\right)\right],
\end{align}
where
\begin{align}
W^{\left(1\right)}=&\frac{\mu_{0}LI^{2}y_{{\rm zp}}}{2\pi d\hbar},\\
W^{\left(2\right)}=&\frac{\mu_{0}LI^{2}y_{{\rm zp}}^{2}}{2\pi d^{2}\hbar}.
\end{align}
For a modest setup~\cite{sazonova2004tunable,ustunel2005modeling,garcia2007mechanical,ning2014transversally,truax2014axially,tsioutsios2017real,liu2019sensing}, $m=1.0\times10^{-22}$~kg, $\omega_{m}=2\pi\times2$~MHz, $L=10$~nm, $d=2$~nm, and $I=380$~nA, we have
\begin{equation}
W^{\left(1\right)}\simeq2\pi\times20\;{\rm kHz},
\end{equation}
which is much smaller than the mechanical resonance frequency $\omega_{m}$, and also have
\begin{equation}
W^{\left(2\right)}\simeq2\pi\times1\;{\rm kHz},
\end{equation}
which is much smaller than the spin-mediated CNT-CNT coupling, for example, $\simeq2\pi\times 12$~kHz, as shown in the section below. Therefore, the direct CNT-CNT coupling can be neglected in our setup.

\section{Controlled Hadamard gate, phase gate, and mechanical quantum delayed-choice experiment}
\label{sec:controlled Hadamard gate}
In order to implement a quantum delayed-choice experiment with macroscopic CNT mechanical resonators, we need a controlled Hadamard gate and a phase gate to act on the CNT mechanical modes. Below, we demonstrate how the effective Hamiltonian in Eq. (\ref{seq:effective-Hamiltonian}) can be used to make all required gates. Let us first consider the controlled Hadamard gate. Tuning the currents to be $I_{k}=I$ and, at the same time, the distances to be $d_{k}=d$ results in a symmetric coupling $g_{k}=g$. The effective Hamiltonian $H_{\text{eff}}$ is accordingly reduced to $H_{\text{eff}}=H_{\text{cnt}}\otimes\sigma_{z}$, where
\begin{equation}\label{seq:beam-splitter-like-Hamiltonian}
H_{\text{cnt}}=\hbar J\left(\sum_{k=1,2}b_{k}^{\dag}b_{k}+b_{1}b_{2}^{\dag}+b_{2}b_{1}^{\dag}\right)
\end{equation}
is a beam-splitter-type interaction, and where
\begin{equation}
J=\frac{2g^{2}\omega_{q}}{\omega_{q}^{2}-\omega_{m}^{2}}
\end{equation}
is an effective CNT-CNT coupling strength. In our discussion, the NV spin is restricted to a subspace spanned by $\left\{|0\rangle,|D\rangle\right\}$, where the spin is a control qubit of a Hadamard gate. The spin in the state $|D\rangle$ mediates the coherent coupling between the separated CNTs, and causes them to evolve under the Hamiltonian $H_{\text{cnt}}$ in Eq.~(\ref{seq:beam-splitter-like-Hamiltonian}). According to the Heisenberg equation of motion, $b_{k}\left(t\right)=\exp\left(iH_{\text{cnt}}t/\hbar\right)b_{k}\exp\left(-iH_{\text{cnt}}t/\hbar\right)$, the unitary evolution for a time $t=\tau_{0}\equiv\pi/\left(4J\right)$ corresponds to a Hadamard-like gate,
\begin{align}
b_{1}\left(\tau_{0}\right)&=\frac{1}{\sqrt{2}}\left(b_{1}-ib_{2}\right),\\
b_{2}\left(\tau_{0}\right)&=\frac{1}{\sqrt{2}}\left(b_{2}-ib_{1}\right).
\end{align}
However, when the spin state is $|0\rangle$, the two CNTs decouple from each other. In this case, their quantum states remain unchanged under the unitary evolution, yielding
\begin{align}
b_{1}\left(t\right)&=b_{1},\\
b_{2}\left(t\right)&=b_{2}.
\end{align}
We have therefore achieved a spin-controlled Hadamard gate between the CNTs. That is, if the NV spin is in the state $|D\rangle$, then the Hadamard operation is applied to the CNTs, and if the NV spin is in the state $|0\rangle$, then the states of the CNTs are unchanged.

We next consider the phase gate. For the phase gate, we tune the currents to be $I_{1}\neq0$ and $I_{2}=0$, such that
$g_{1}=g$ and $g_{2}=0$, causing the effective Hamiltonian in Eq.~(\ref{seq:effective-Hamiltonian}) to become
\begin{equation}\label{effective-Hamiltonian-for-phase-gate}
H_{\text{cnt}}=\hbar Jb_{1}^{\dag}b_{1}\sigma_{z}.
\end{equation}
We find from Eq. (\ref{effective-Hamiltonian-for-phase-gate}) that there exists a spin-induced shift, $J$, of the mechanical resonance. This dispersive shift can, in turn, introduce a dynamical phase, $\phi\left(t\right)=Jt$, onto the first CNT. With the spin being in the state $|D\rangle$, we solve the Heisenberg equations of motion for the CNTs, and then obtain a phase gate,
\begin{align}
b_{1}\left(t\right)&=\exp\left[i\phi\left(t\right)\right]b_{1},\\
b_{2}\left(t\right)&=b_{2}.
\end{align}
In fact, similar to the controlled Hadamard gate discussed above, the phase gate can also be controlled by the spin according to Eq. (\ref{effective-Hamiltonian-for-phase-gate}).

Having achieved all required gates, we now turn to the detailed description of the macroscopic quantum delayed-choice experiment with CNT resonators. The hybrid system is initially prepared in the state $|\Psi\rangle_{i}\equiv|\Psi\left(0\right)\rangle=\left(b_{1}^{\dag}\otimes \mathcal{I}_{2}|\text{vac}\rangle\right)\otimes|D\rangle$, where $|\text{vac}\rangle$ refers to the phonon vacuum of the CNTs and $\mathcal{I}_{2}$ is the identity operator on the second CNT. First, we turn on the currents of the CNTs and ensure $I_{k}=I$. After a time $\tau_{0}$, a Hadamard operation is applied to the CNTs and accordingly, $|\Psi\rangle_{i}$ becomes
\begin{equation}
|\Psi\left(\tau_{0}\right)\rangle=\frac{1}{\sqrt{2}}\left(b_{1}^{\dag}
+ib_{2}^{\dag}\right)|\text{vac}\rangle|D\rangle.
\end{equation}
Then, we turn off the current of the second CNT for a phase accumulation for a time $\tau_{1}$. As a consequence, the system further evolves to
\begin{equation}
|\Psi\left(\tau_{0}+\tau_{1}\right)\rangle
=\frac{1}{\sqrt{2}}\left[\exp\left(i\phi\right)b_{1}^{\dag}+ib_{2}^{\dag}\right]|\text{vac}\rangle|D\rangle.
\end{equation}
While achieving the desired phase $\phi$, we make a spin single-qubit rotation $|D\rangle\rightarrow\cos\left(\varphi\right)|0\rangle+\sin\left(\varphi\right)|D\rangle$, and have \begin{align}\label{seq:state-before-second-BS}
|\Psi\left(\tau_{0}+\tau_{1}\right)\rangle=\frac{1}{\sqrt{2}}\left[\exp\left(i\phi\right)b_{1}^{\dag}
+ib_{2}^{\dag}\right]|\text{vac}\rangle\left(\cos\varphi|0\rangle+\sin\varphi|D\rangle\right).
\end{align}
Here, note that, we have ignored the length of the driving pulse of the spin rotation as being of the order of~ns, and thus assumed that the state of the CNTs remains unchanged. At the end of the driving pulse, we turn on the current of the second CNT again and hold for another $\tau_{0}$ to perform a Hadamard gate. This gate is in a quantum superposition of being present and absent. The three operations on the mechanical phonon correspond to the actions, on a single photon, of the input beam splitter, the phase shifter, and the output beam splitter, respectively, in quantum delayed-choice experiments with a Mach-Zehnder interferometer. The final state is therefore given by
\begin{align}\label{seq:finial-quantum-state}
|\Psi\rangle_{f}\equiv|\Psi\left(2\tau_{0}+\tau_{1}\right)\rangle=\cos\left(\varphi\right)|\text{particle}\rangle|0\rangle
+\sin\left(\varphi\right)|\text{wave}\rangle|D\rangle,
\end{align}
\begin{figure}[tbph]
	\centering
	\includegraphics[width=13.0cm]{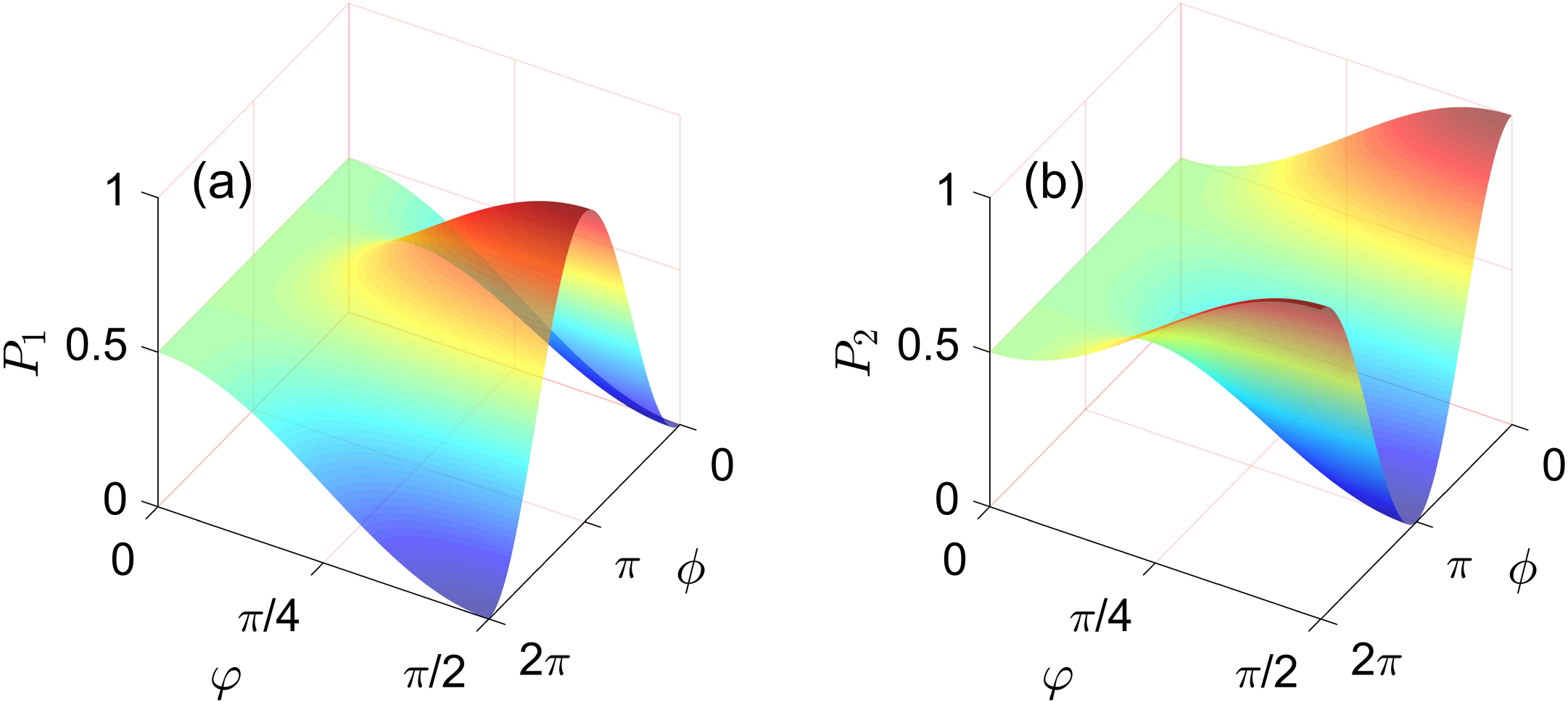}
	\caption{(Color online) (a) Probability $P_{1}$ and (b) $P_{2}$ as a function of the rotation angle $\varphi$ and the relative phase $\phi$. This represents a continuous transition between a particle-type behavior ($\varphi=0$) and a wave-type behavior ($\varphi=\pi/2$).}\label{sfig:ideal-morphing}
\end{figure}
where
\begin{align}
|\text{particle}\rangle&=\frac{1}{\sqrt{2}}\left[\exp\left(i\phi\right)b_{1}^{\dag}
+ib_{2}^{\dag}\right]|\text{vac}\rangle,\\
|\text{wave}\rangle&=\frac{1}{2}\left\{\left[\exp\left(i\phi\right)-1\right]b_{1}^{\dag}
+i\left[\exp\left(i\phi\right)+1\right]b_{2}^{\dag}\right\}|\text{vac}\rangle,
\end{align}
describe particle and wave behaviors, respectively. This reveals that the CNT mechanical phonon is in a quantum superposition of both a particle and a wave. The probability of finding a single phonon in the $k$th CNT is expressed as
\begin{equation}\label{seq:detection-probablity}
P_{k}=\frac{1}{2}+(-1)^{k}\frac{1}{2}\sin^{2}\left(\varphi\right)\cos\left(\phi\right),
\end{equation}
according to Eq.~(\ref{seq:finial-quantum-state}). In Fig. \ref{sfig:ideal-morphing}, we have plotted the probabilities $P_{1}$ and $P_{2}$ versus the rotation angle $\varphi$ and the relative phase $\phi$.  In this figure we find that the mechanical phonon shows a morphing behavior between particle ($\varphi=0$) and wave ($\varphi=\pi/2$).

\begin{figure}[t]
	\centering
	\includegraphics[width=9.0cm]{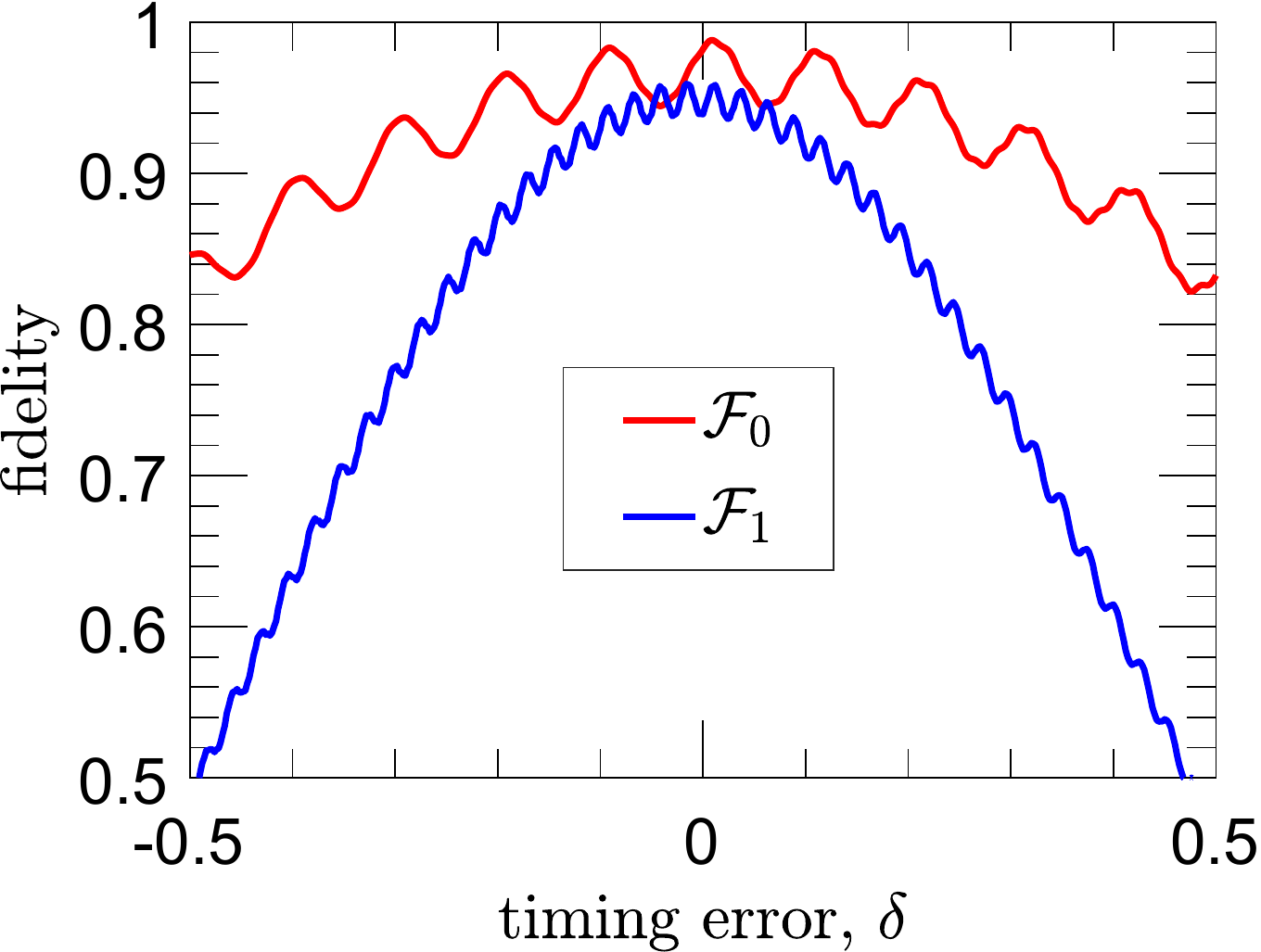}
	\caption{(Color online) Gate fidelity as a function of the timing error. We have assumed that $\delta_{0}/\tau_{0}=\delta_{1}/\tau_{1}=\delta$,
		and that the initial state is: (i) $b^{\dag}_{1}|{\rm vac}\rangle|D\rangle$
		for the Hadamard gate, and (ii) $\frac{1}{\sqrt{2}}\left(b_{1}^{\dag}+ib_{2}^{\dag}
		\right)|{\rm vac}\rangle|D\rangle$ for the phase gate that accumulates a
		relative phase $\pi$. Here, in addition to
		$\gamma_{s}/2\pi=200\gamma_{m}/2\pi=80$~Hz, we have assumed that
		$g/2\pi=100$~kHz, $\omega_{m}/2\pi=2$~MHz, $\Omega=10\omega_{m}$,
		and $\Delta_{-}=142\omega_{m}$, resulting in
		$\omega_{q}\simeq1.5\omega_{m}$ and then $J/2\pi\simeq12$~kHz. We
		have also assumed that $n_{\text{th}}=100$, which corresponds to
		the environment temperature of $\simeq10$~mK.}\label{sfig_timing_error}
\end{figure}

\vspace{3mm}

\noindent
We now consider the timing errors of the Hadamard and phase gates. We first consider the Hadamard gate. We
assume that the error of the time required for performing the
Hadamard gate is $\delta_{0}$, such that the actual evolution time
for the gate becomes $\tau_{0}^{\prime}=\tau_{0}+\delta_{0}$. In
order to estimate the effect of this timing error on the gate
performance, we introduce a gate fidelity, defined as
\begin{align}
\mathcal{F}_{0}=\langle\Psi_{{\rm target},0}|\rho_{{\rm
		actual},0}\left(\tau_{0}^{\prime}\right)|\Psi_{{\rm
		target},0}\rangle,
\end{align}
where $|\Psi_{{\rm target}, 0}\rangle$ is the target state given
by the ideal Hadamard gate, and $\rho_{{\rm
		actual},0}\left(\tau_{0}^{\prime}\right)$ is the actual state
obtained by integrating the exact master equation, given by
Eq.~(\ref{seq:exact-full-master-equation}). In
Fig.~\ref{sfig_timing_error}, we plot the gate fidelity,
$F_{0}$, versus the timing error $\delta_{0}$ (red
curve). In this figure, the initial state for the Hadamard gate is
assumed to be $b^{\dag}_{1}|{\rm vac}\rangle|D\rangle$, so that
the target state is $|\Psi_{{\rm target},
	0}\rangle=\frac{1}{\sqrt{2}}\left(b_{1}^{\dag}+ib_{2}^{\dag}\right)|{\rm
	vac}\rangle|D\rangle$. Here, $|{\rm vac}\rangle$ represents the
acoustic vacuum state of the CNT resonators. From this
figure, we find that for
$-0.32\tau_{0}\lesssim\delta_{0}\lesssim0.34\tau_{0}$, the gate
fidelity $\mathcal{F}_{0}$ can be kept above 0.9.

For the phase gate, we assume, as above, that the timing error is
$\delta_{1}$. Thus, the actual evolution time for the phase gate
is $\tau_{1}^{\prime}=\tau_{1}+\delta_{1}$. We also introduce a
gate fidelity, defined as
\begin{align}
\mathcal{F}_{1}=\langle\Psi_{{\rm target},1}|\rho_{{\rm
		actual},1}\left(\tau_{1}^{\prime}\right)|\Psi_{{\rm
		target},1}\rangle,
\end{align}
where $|\Psi_{{\rm target}, 1}\rangle$ is the target state given
by the ideal phase gate, and $\rho_{{\rm
		actual},1}\left(\tau_{1}^{\prime}\right)$ is the actual state
obtained from the exact master equation, given in Eq.~(\ref{seq:exact-full-master-equation}). The gate fidelity
$\mathcal{F}_{1}$ is plotted as a function of the timing error
$\delta_{1}$ in Fig.~(\ref{sfig_timing_error}) (blue curve).
There, we assumed that the initial state for the phase gate is
$\frac{1}{\sqrt{2}}\left(b_{1}^{\dag}+ib_{2}^{\dag}\right)$, and
that the phase accumulated is equal to $\pi$. The target state
$|\Psi_{{\rm target}, 1}\rangle$ is, therefore, given by
$\frac{1}{\sqrt{2}}\left(-b_{1}^{\dag}+ib_{2}^{\dag}\right)$. It
is seen from this figure that, as long as
$-0.15\tau_{1}\lesssim\delta_{1}\lesssim0.12\tau_{1}$, we can
obtain the gate fidelity of $\mathcal{F}_{1}>0.9$.

\begin{figure}[tbph]
	\centering
	\includegraphics[width=9.0cm]{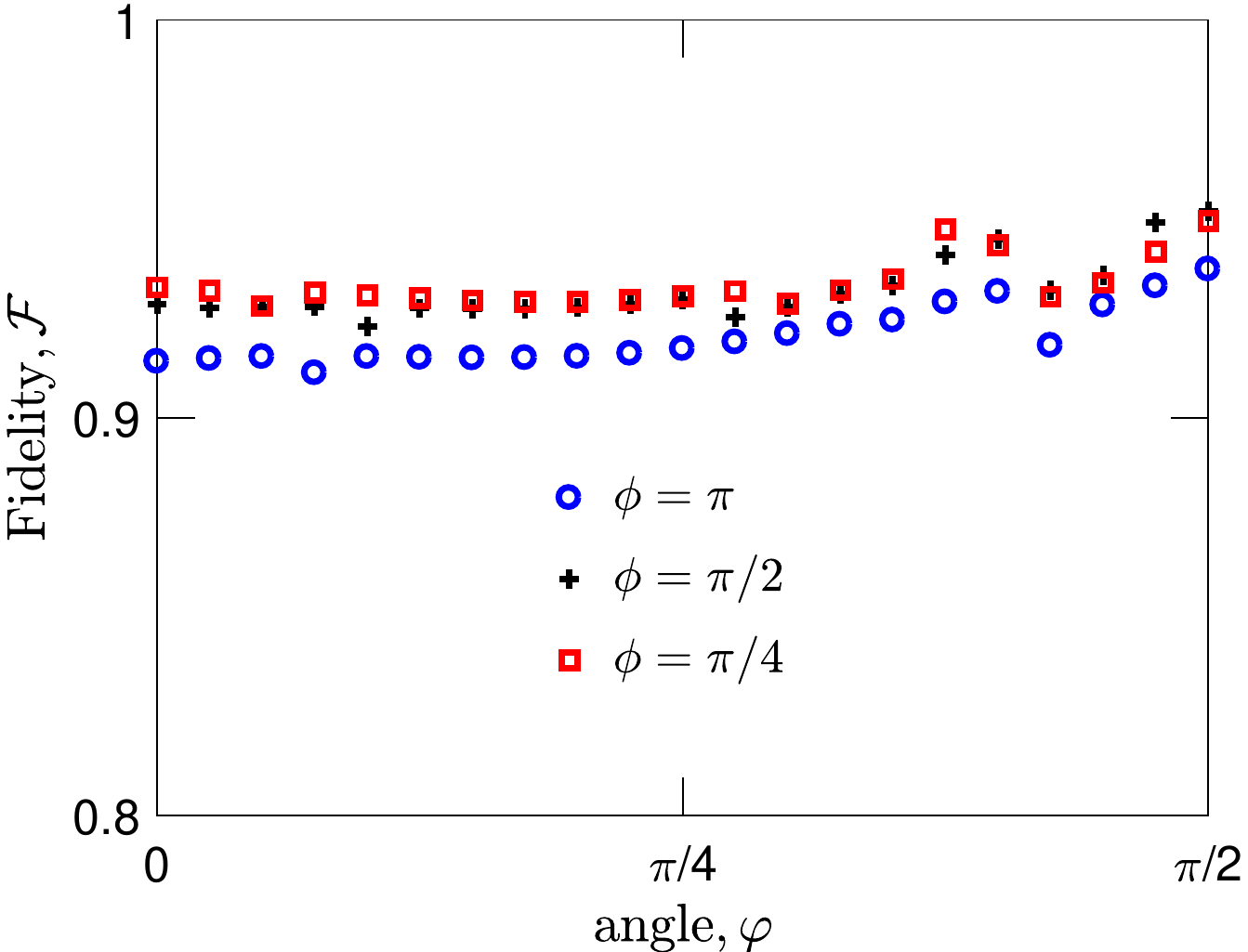}
	\caption{(Color online) (a) Fidelity $\mathcal{F}$ as a function
		of the rotation angle $\varphi$. All the results are numerically
		obtained by integrating the exact master equation in
		Eq.~(\ref{seq:exact-full-master-equation}). Here, in addition to
		$\gamma_{s}/2\pi=200\gamma_{m}/2\pi=80$~Hz, we have assumed that
		$g/2\pi=100$~kHz, $\omega_{m}/2\pi=2$~MHz, $\Omega=10\omega_{m}$,
		and $\Delta_{-}=142\omega_{m}$, resulting in
		$\omega_{q}\simeq1.5\omega_{m}$ and then $J/2\pi\simeq12$~kHz. We
		have also assumed that $n_{\text{th}}=100$, which corresponds to
		an environmental temperature of $\simeq
		10$~mK.}\label{sfig:s-fidelity}
\end{figure}

Note that the spin, in a classical mixed state of the form $\cos^{2}\left(\varphi\right)|0\rangle\langle0|+\sin^{2}\left(\varphi\right)|D\rangle\langle D|$, would lead to the same measured statistics in Eq.~(\ref{seq:detection-probablity}), that is, a local hidden variable model is capable of reproducing the quantum predictions. This is a loophole~\cite{chaves2018causal,huang2018loophole,polino2018device,yu2018experimental}. However, as discussed in Refs.~\cite{peruzzo2012quantum,kaiser2012entanglement,zheng2015quantum,liu2017twofold}, this loophole can be avoided as long as the second Hadamard operation is ensured to be in a truly quantum superposition of being present and absent. In our proposal, the second Hadamard operation is conditioned on the spin state. If the spin is in the $\ket{0}$ state, then the Hadamard operation is absent; if the spin is in the $\ket{D}$ state, then the Hadamard operation is present; if the spin is in a quantum superposition of the $\ket{0}$ and $\ket{D}$ states, then the Hadamard operation is in a quantum superposition of being present and absent. To confirm such a quantum superposition, in Fig.~(\ref{sfig:s-fidelity}) we numerically calculate the fidelity, $\mathcal{F}=\leftidx{_{f}}\langle\Psi|\rho_{\rm actual}\left(\tau_{T}\right)|\Psi\rangle_{f}$, between the desired state $|\Psi\rangle_{f}$ in Eq.~(\ref{seq:finial-quantum-state}) and the actual state $\rho_{\rm actual}\left(\tau_{T}\right)$ obtained from the exact master equation in Eq.~(\ref{seq:exact-full-master-equation}). From this figure, we find that the fidelity is very close to unity even for the finite temperature of $T\simeq10$~mK. Furthermore, in experiments, in order to exclude the classical interpretation and prove the existence of the coherent wave-particle superposition, the quantum coherence between the states $|0\rangle$ and $|D\rangle$ should be verified. Experimentally, this coherence can be prepared by a spin single-qubit operation~\cite{huang2011observation,
	lillie2017environmentally,xing2017experimental}, and can be verified by performing quantum state tomography to show all the elements of the density matrix of the spin~\cite{xing2017experimental}.

\section{Phonon occupation at finite temperatures}
\label{sec:Phonon occupations at finite temperature}
We begin by considering the total operation time, which is given by $\tau_{T}=2\tau_{0}+\tau_{1}$, as discussed in Sec. \ref{sec:controlled Hadamard gate}. Here, $\tau_{0}=\pi/\left(4J\right)$ is the time for the Hadamard gate and $\tau_{1}\in\left[0,2\pi/J\right]$ is the time for the phase gate. In a realistic setup, we can assume $\omega_{m}/2\pi\simeq 2$~MHz, $\omega_{q}/2\pi\simeq3$~MHz, and $g/2\pi=100$~kHz, such that $J/2\pi\simeq12$~kHz, yielding a maximum total time $\tau_{T}^{\text{max}}=2\tau_{0}+\tau_{1}^{\text{max}}\simeq0.1$~ms, where $\tau_{1}^{\text{max}}=2\pi/J$ is the maximum phase gate time. Note that, the operation time $\tau_{T}$ depends inversely on the CNT-CNT coupling strength $J$, but the enhancement in $J$ is limited by the validity of the effective Hamiltonian $H_{{\rm eff}}$.

The total decoherence in our setup can be divided into two parts, one from the spin and the other from the CNTs. The spin decoherence in general includes the relaxation and the dephasing. For an NV electronic spin, the relaxation time $T_{1}$ can reach up to several minutes at low temperatures and the dephasing time can be $T_{2}\simeq 2$~ms even at room temperature~\cite{balasubramanian2009ultralong,bar2013solid}. These justify neglecting the spin decoherence. For the mechanical decoherence, despite a long phonon life, the low mechanical frequency makes the CNT mechanical modes very sensitive to the environmental temperature. In this section and in Sec. \ref{sec:Signal-to-noise ratio at finite temperature}, we discuss the effects of the mechanical noise on our quantum delayed-choice experiment, and demonstrate that the morphing between wave and particle can still be effectively observed even at finite temperatures.

As a result, the dissipative processes, in the hybrid system considered here, are induced only by the mechanical decoherence, which arises from the vacuum fluctuation and thermal noise. The full dynamics of the system can then be governed by the following master equation
\begin{align}\label{seq:full-master-equation}
\dot{\rho}\left(t\right)=\frac{i}{\hbar}\left[\rho\left(t\right),H\left(t\right)\right]-\frac{\gamma_{m}}{2}n_{\text{th}}
\sum_{k=1,2}\mathcal{L}\left(b_{k}^{\dag}\right)\rho\left(t\right)
-\frac{\gamma_{m}}{2}\left(n_{\text{th}}+1\right)\sum_{k=1,2}\mathcal{L}\left(b_{k}\right)\rho\left(t\right),
\end{align}
where $\rho$ is the density operator of the system, $\gamma_{m}$ is the mechanical decay rate, $n_{\text{th}}=\left[\exp\left(\hbar\omega_{m}/k_{B}T\right)-1\right]^{-1}$ is the equilibrium phonon occupation at temperature $T$, and $\mathcal{L}\left(o\right)\rho\left(t\right)=o^{\dag}o\rho\left(t\right)-2o\rho\left(t\right)o^{\dag}+\rho\left(t\right)o^{\dag}o$ is the Lindblad superoperator. Here, $H\left(t\right)$ is a binary Hamiltonian of the form,
\begin{align}\label{seq:ternary-Hamiltonian}
H\left(t\right)=
\begin{cases}
H_{0}, & 0<t\leq\tau_{0},\; {\rm and}\;\; \tau_{0}+\tau_{1}<t\leq\tau_{T}\\
H_{1}, & \tau_{0}<t\leq\tau_{0}+\tau_{1},\\
\end{cases}
\end{align}
with
\begin{align}
H_{0}= &\hbar J\left(\sum_{k=1,2}b_{k}^{\dag}b_{k}+b_{1}b_{2}^{\dag}+b_{2}b_{1}^{\dag}\right)\sigma_{z},\\
H_{1}= &\hbar Jb_{1}^{\dag}b_{1}\sigma_{z}.
\end{align}
The three time intervals in Eq.~(\ref{seq:ternary-Hamiltonian}) correspond to the first Hadamard gate, the phase gate and the second Hadamard gate, respectively. Note that in Eq.~(\ref{seq:ternary-Hamiltonian}), we did not include the spin single-qubit rotation before the third interval because the length of the driving pulse is of the order of~ns. We can derive the system evolution step by step.

Let us now consider the first evolution interval $0<t\leq\tau_{0}$. During this interval, the coupling of the CNT mechanical modes introduces two delocalized phononic modes,
\begin{equation}\label{seq:delocalized-phononic-modes}
c_{\pm}=\frac{1}{\sqrt{2}}\left(b_{1}\pm b_{2}\right),
\end{equation}
such that $H_{0}$ is diagonalized to be
\begin{equation}
H_{0}=2\hbar Jc_{+}^{\dag}c_{+}\sigma_{z},
\end{equation}
and the master equation in Eq.~(\ref{seq:full-master-equation}) is reexpressed, in terms of the modes $c_{\pm}$, as
\begin{align}\label{seq:delocalized-mode-master-equation}
\dot{\rho}=i\left[\rho,2Jc_{+}^{\dag}c_{+}\sigma_{z}\right]-\frac{\gamma_{m}}{2}n_{\text{th}}\sum_{\mu=1,2}
\mathcal{L}\left(c_{\mu}^{\dag}\right)\rho\left(t\right)
-\frac{\gamma_{m}}{2}\left(n_{\text{th}}+1\right)\sum_{\mu=1,2}L\left(c_{\mu}\right)\rho\left(t\right).
\end{align}
In order to calculate the phonon occupations at the end of the first interval, we need to obtain the equations of motion for $\langle c_{\pm}^{\dag}c_{\pm}\rangle$, $\langle c_{+}^{\dag}c_{-}\rangle$, $\langle c_{+}^{\dag}c_{-}\sigma_{z}\rangle$, and $\langle c_{+}^{\dag}c_{-}\sigma_{z}^{2}\rangle$. Here, $\langle O\rangle$ represents the expectation value of the operator $O$. Following the master equation in Eq.~(\ref{seq:delocalized-mode-master-equation}), we have
\begin{align}
\frac{d}{dt}\langle c_{\pm}^{\dag}c_{\pm}\rangle=&-\gamma_{m}\langle c_{\pm}^{\dag}c_{\pm}\rangle+\gamma_{m}n_{\text{th}},\label{seq:equation-of-motion-for-delocalized-mode-occupations}\\
\frac{d}{dt}\langle c_{+}^{\dag}c_{-}\rangle=&i2J\langle c_{+}^{\dag}c_{-}\sigma_{z}\rangle-\gamma_{m}\langle c_{+}^{\dag}c_{-}\rangle,\label{seq:equation-of-motion-for-delocalized-mode-correlation}\\
\frac{d}{dt}\langle c_{+}^{\dag}c_{-}\sigma_{z}\rangle=&i2J\langle c_{+}^{\dag}c_{-}\sigma_{z}^{2}\rangle-\gamma_{m}\langle c_{+}^{\dag}c_{-}\sigma_{z}\rangle,\label{seq:equation-of-motion-for-delocalized-mode-spin-correlation-1}\\
\frac{d}{dt}\langle c_{+}^{\dag}c_{-}\sigma_{z}^{2}\rangle=&i2J\langle c_{+}^{\dag}c_{-}\sigma_{z}\rangle-\gamma_{m}\langle c_{+}^{\dag}c_{-}\sigma_{z}^{2}\rangle,\label{seq:equation-of-motion-for-delocalized-mode-spin-correlation-2}
\end{align}
where we have used the relation $\sigma_{z}^{3}=\sigma_{z}$. We can straightforwardly solve the differential equation~(\ref{seq:equation-of-motion-for-delocalized-mode-occupations}) to find
\begin{equation}\label{seq:delocalized_mode_occupation_in_first_interval}
\langle c_{\pm}^{\dag}c_{\pm}\rangle\left(t\right)=\left(\frac{1}{2}-n_{\text{th}}\right)\exp\left(-\gamma_{m}t\right)
+n_{\text{th}}.
\end{equation}
Combining Eqs. (\ref{seq:equation-of-motion-for-delocalized-mode-spin-correlation-1}) and (\ref{seq:equation-of-motion-for-delocalized-mode-spin-correlation-2}) gives
\begin{align}\label{seq:cplusdagcminus_z}
\langle c_{+}^{\dag}c_{-}\sigma_{z}^{j}\rangle\left(t\right)=\left(-1\right)^{j}\frac{1}{2}
\exp\left(-i2Jt\right)\exp\left(-\gamma_{m}t\right),
\end{align}
for $j=1,2$. Upon substituting Eq. (\ref{seq:cplusdagcminus_z}) back into Eq.~(\ref{seq:equation-of-motion-for-delocalized-mode-correlation}), we can then obtain
\begin{equation}
\langle c_{+}^{\dag}c_{-}\rangle\left(t\right)=\frac{1}{2}\exp\left(-i2Jt\right)\exp\left(-\gamma_{m}t\right).
\end{equation}
It is found, according to Eq.~(\ref{seq:delocalized-phononic-modes}), that in the localized-mode basis, \begin{align}\label{CNT-occupation-right-after-first-interval1}
\langle b_{k}^{\dag}b_{k}\rangle\left(\tau_{0}\right)&=\left(\frac{1}{2}-n_{\text{th}}\right)
\exp\left(-\gamma_{m}\tau_{0}\right)+n_{\text{th}},\\
\label{CNT-occupation-right-after-first-interval2}
\langle b_{1}^{\dag}b_{2}\rangle\left(\tau_{0}\right)&=\frac{i}{2}\exp\left(-\gamma_{m}\tau_{0}\right).
\end{align}

For the second evolution interval $\tau_{0}<t\leq\tau_{0}+\tau_{1}$, we directly use the master equation in Eq.~(\ref{seq:full-master-equation}) but with $H\left(t\right)$ replaced by $H_{1}$. When comparing with the master equation in Eq.~(\ref{seq:delocalized-mode-master-equation}), we see that the equations of motion for $\langle b_{k}^{\dag}b_{k}\rangle$, $\langle b_{1}^{\dag}b_{2}\rangle$, $\langle b_{1}^{\dag}b_{2}\sigma_{z}\rangle$, and $\langle b_{1}^{\dag}b_{2}\sigma_{z}^{2}\rangle$ should have the same forms as in Eqs.~(\ref{seq:equation-of-motion-for-delocalized-mode-occupations}), (\ref{seq:equation-of-motion-for-delocalized-mode-correlation}), (\ref{seq:equation-of-motion-for-delocalized-mode-spin-correlation-1}), and (\ref{seq:equation-of-motion-for-delocalized-mode-spin-correlation-2}), but with the substitutions $c_{+}\rightarrow b_{1}$, $c_{-}\rightarrow b_{2}$ and $2J\rightarrow J$. In combination with the initial conditions, given in Eqs.~(\ref{CNT-occupation-right-after-first-interval1}) and (\ref{CNT-occupation-right-after-first-interval2}), we follow the same procedure as above to find
\begin{align}
\langle b_{k}^{\dag}b_{k}\rangle\left(\tau_{0}+\tau_{1}\right)
=&\left(\frac{1}{2}-n_{\text{th}}\right)\exp\left[-\gamma_{m}\left(\tau_{0}+\tau_{1}\right)\right]+n_{\text{th}},
\label{CNT-occupation-right-after-second-interval}\\
\langle b_{1}^{\dag}b_{2}\rangle\left(\tau_{0}+\tau_{1}\right)=&\frac{i}{2}\exp\left(-iJ\tau_{1}\right)
\exp\left[-\gamma_{m}\left(\tau_{0}+\tau_{1}\right)\right].
\label{CNT-correlation-right-after-second-interval}
\end{align}

We now turn to the third evolution interval $\tau_{0}+\tau_{1}<t\leq2\tau_{0}+\tau_{1}$. Before this interval or
at the end of the second interval, we apply a single qubit rotation, $|D\rangle\rightarrow\cos\left(\varphi\right)|0\rangle+\sin\left(\varphi\right)|D\rangle$, on the NV spin to engineer the subsequent Hadamard operation to be in a quantum superposition of being absent and present. In this situation, we still use the delocalized-mode basis and the corresponding master equation in Eq.~(\ref{seq:delocalized-mode-master-equation}). According to Eqs.~(\ref{CNT-occupation-right-after-second-interval}) and (\ref{CNT-correlation-right-after-second-interval}), the initial conditions of the last evolution can be rewritten, in terms of $c_{\pm}$, as
\begin{align}
\langle c_{\pm}^{\dag}c_{\pm}\rangle\left(\tau_{0}+\tau_{1}\right)=&\left[\frac{1}{2}\pm\frac{1}{2}\sin\left(J\tau_{1}\right)-n_{\text{th}}\right]
\exp\left[-\gamma\left(\tau_{0}+\tau_{1}\right)\right]+n_{\text{th}},\\
\langle c_{+}^{\dag}c_{-}\rangle\left(\tau_{0}+\tau_{1}\right)=&-\frac{i}{2}\cos\left(J\tau_{1}\right)
\exp\left[-\gamma\left(\tau_{0}+\tau_{1}\right)\right],\\
\langle c_{+}^{\dag}c_{-}\sigma_{z}^{j}\rangle\left(\tau_{0}+\tau_{1}\right)=&\left(-1\right)^{j}\sin^{2}\left(\varphi\right)
\langle c_{+}^{\dag}c_{-}\rangle\left(\tau_{0}+\tau_{1}\right),
\end{align}
for $j=1,2$. Then, as before, solving the differential equations in Eqs.~(\ref{seq:equation-of-motion-for-delocalized-mode-occupations}), (\ref{seq:equation-of-motion-for-delocalized-mode-correlation}), (\ref{seq:equation-of-motion-for-delocalized-mode-spin-correlation-1}) and (\ref{seq:equation-of-motion-for-delocalized-mode-spin-correlation-2}) leads to
\begin{align}
\langle c_{\pm}^{\dag}c_{\pm}\rangle\left(t\right)=&\left[\frac{1}{2}\pm\frac{1}{2}\sin\left(J\tau_{1}\right)-n_{\text{th}}\right]
\exp\left(-\gamma_{m}t\right)+n_{\text{th}},\\
\langle c_{+}^{\dag}c_{-}\rangle\left(t\right)=&-\frac{i}{2}\cos\left(J\tau_{1}\right)
\left\{\cos^{2}\left(\varphi\right)+\sin^{2}\left(\varphi\right)\exp\left[-i2J\left(t-\tau_{0}-\tau_{1}\right)\right]\right\}
\exp\left(-\gamma_{m}t\right),
\end{align}
which, in turn, gives
\begin{equation}
n_{k}\equiv\langle b_{k}^{\dag}b_{k}\rangle\left(\tau_{T}\right)
=\left(P_{k}-n_{\text{th}}\right)
\exp\left(-\gamma_{m}\tau_{T}\right)+n_{\text{th}},
\end{equation}
which is the phonon occupation of the $k$th at the end of the third interval. For a realistic CNT, the mechanical linewidth can be set to $\gamma_{m}/2\pi=0.4$~Hz~\cite{moser2014nanotube}, and then we obtain a phonon lifetime of $\simeq400$~ms, which is much longer than the maximum total time $\tau_{T}^{{\rm max}}\simeq0.1$~ms. This ensures $\gamma_{m}\tau_{T}\ll1$, which results in
\begin{equation}\label{seq:finial_occupations_with_mechanical_noises}
n_{k}\simeq P_{k}+n_{\text{th}}\gamma_{m}\tau_{T}.
\end{equation}
This shows that the occupation for each CNT has two contributions: one from a coherent phonon signal and one from thermal excitations. Furthermore, we find from Eq.~(\ref{seq:finial_occupations_with_mechanical_noises}) that the thermal excitations have equal contributions to $n_{1}$ and $n_{2}$. This is because the thermal excitations do not contribute to the interference. For an environmental temperature $T=10$~mK, the equilibrium phonon occupation is $n_{{\rm th}}\simeq100$, yielding  $n_{\text{th}}\gamma_{m}\tau_{T}^{{\rm max}}\simeq0.03$, which can be neglected, as shown in Fig. 2 of the article.

\section{Signal-to-noise ratio at finite temperatures}
\label{sec:Signal-to-noise ratio at finite temperature}
In addition to the thermal occupation, $n_{\text{th}}\gamma_{m}\tau_{T}$, in Eq.~(\ref{seq:finial_occupations_with_mechanical_noises}), the desired signal $P_{k}$ is also always accompanied by fluctuation noise. Such a noise includes vacuum fluctuations and thermal fluctuations. In particular, the latter increases with temperature, so that the signal can be completely drowned in the noise when the temperature is sufficiently high. In this case, it is very difficult to observe the morphing between wave and particle. Thus in this section, we analyze this fluctuation noise in detail, and demonstrate that, in order for the morphing behavior to be observed effectively, the total fluctuation noise of both CNTs should be limited by an upper bound, which leads to a critical temperature $T_{\rm c}$.

Specifically, we begin by deriving the fluctuation $\delta n_{k}$ in the occupation $n_{k}$, for $k=1,2$. This is defined by
\begin{align}\label{seq:phonon_fluctuation}
\left(\delta n_{k}\right)^{2}&=\langle \left(b_{k}^{\dag}b_{k}\right)^{2}\rangle\left(\tau_{T}\right)-\langle b_{k}^{\dag}b_{k}\rangle^2\left(\tau_{T}\right)\nonumber\\
&=\langle b_{k}^{\dag}b_{k}^{\dag}b_{k}b_{k}\rangle\left(\tau_{T}\right)+n_{k}-n_{k}^2.
\end{align}
In order to understand the fluctuation noise better, we need to derive an analytical expression of $\delta n_{k}$. In Sec.~\ref{sec:Phonon occupations at finite temperature}, $n_{k}$ has been given in Eq.~(\ref{seq:finial_occupations_with_mechanical_noises}). Below, we derive the evolution of $\langle b_{k}^{\dag}b_{k}^{\dag}b_{k}b_{k}\rangle$ in a step-by-step manner as in Sec.~\ref{sec:Phonon occupations at finite temperature}.

We now consider the first evolution interval $0<t\leq\tau_{0}$. During this interval, the delocalized modes $c_{\pm}$ in Eq.~(\ref{seq:delocalized-phononic-modes}) are employed owing to the coupling of the CNT mechanical modes, and the dynamics is described by the master equation in Eq.~(\ref{seq:delocalized-mode-master-equation}). To achieve $\langle b_{k}^{\dag}b_{k}^{\dag}b_{k}b_{k}\rangle$ at time $\tau_{T}$, the dynamical evolutions of $\langle c_{\pm}^{\dag}c_{\pm}^{\dag}c_{\pm}c_{\pm}\rangle$, $\langle c_{+}^{\dag}c_{+}c_{-}^{\dag}c_{-}\rangle$, $\langle c_{+}^{\dag}c_{+}^{\dag}c_{+}c_{-}\rangle$, $\langle c_{+}^{\dag}c_{-}^{\dag}c_{-}c_{-}\rangle$, and $\langle c_{+}^{\dag}c_{+}^{\dag}c_{-}c_{-}\rangle$ are involved. The equations of motion for $\langle c_{\pm}^{\dag}c_{\pm}^{\dag}c_{\pm}c_{\pm}\rangle$ and $\langle c_{+}^{\dag}c_{+}c_{-}^{\dag}c_{-}\rangle$ are
\begin{align}
\frac{d}{dt}\langle c_{\pm}^{\dag}c_{\pm}^{\dag}c_{\pm}c_{\pm}\rangle&=4\gamma_{m}n_{\text{th}}\langle c_{\pm}^{\dag}c_{\pm}\rangle-2\gamma_{m}\langle c_{\pm}^{\dag}c_{\pm}^{\dag}c_{\pm}c_{\pm}\rangle,\\
\frac{d}{dt}\langle c_{+}^{\dag}c_{+}c_{-}^{\dag}c_{-}\rangle&=\gamma_{m}n_{\text{th}}\left(\langle c_{+}^{\dag}c_{+}\rangle+\langle c_{-}^{\dag}c_{-}\rangle\right)-2\gamma_{m}\langle c_{+}^{\dag}c_{+}c_{-}^{\dag}c_{-}\rangle.
\end{align}
Substituting Eq. (\ref{seq:delocalized_mode_occupation_in_first_interval}) yields
\begin{align}
\label{seq:first_interval_cpmdagcpmcdagcpmcpm}
\langle c_{\pm}^{\dag}c_{\pm}^{\dag}c_{\pm}c_{\pm}\rangle\left(\tau_{0}\right)&=2X\left(\tau_{0}\right),\\
\label{seq:first_intrval_cplusdagcpluscminusdagcminus}
\langle c_{+}^{\dag}c_{+}c_{-}^{\dag}c_{-}\rangle\left(\tau_{0}\right)&=X\left(\tau_{0}\right),
\end{align}
where
\begin{equation}
X\left(t\right)=n_{\text{th}}\left(n_{\text{th}}-1\right)\exp\left(-2\gamma_{m}t\right)
+n_{\text{th}}\left(1-2n_{\text{th}}\right)\exp\left(-\gamma_{m}t\right)+n_{\text{th}}^{2}.
\end{equation}
The equations of motion for $\langle c_{+}^{\dag}c_{+}^{\dag}c_{+}c_{-}\rangle$ are found to be
\begin{align}
\label{seq:cplusdagcplusdagcpluscminus}
\frac{d}{dt}\langle c_{+}^{\dag}c_{+}^{\dag}c_{+}c_{-}\rangle&=i2J\langle c_{+}^{\dag}c_{+}^{\dag}c_{+}c_{-}\sigma_{z}\rangle+2\gamma_{m}n_{\text{th}}\langle c_{+}^{\dag}c_{-}\rangle-2\gamma_{m}\langle c_{+}^{\dag}c_{+}^{\dag}c_{+}c_{-}\rangle,\\
\label{seq:cplusdagcplusdagcpluscminussigmaz}
\frac{d}{dt}\langle c_{+}^{\dag}c_{+}^{\dag}c_{+}c_{-}\sigma_{z}\rangle&=i2J\langle c_{+}^{\dag}c_{+}^{\dag}c_{+}c_{-}\sigma_{z}^{2}\rangle+2\gamma_{m}n_{\text{th}}\langle c_{+}^{\dag}c_{-}\sigma_{z}\rangle-2\gamma_{m}\langle c_{+}^{\dag}c_{+}^{\dag}c_{+}c_{-}\sigma_{z}\rangle,\\
\label{seq:cplusdagcplusdagcpluscminussigmaz2}
\frac{d}{dt}\langle c_{+}^{\dag}c_{+}^{\dag}c_{+}c_{-}\sigma_{z}^{2}\rangle&=i2J\langle c_{+}^{\dag}c_{+}^{\dag}c_{+}c_{-}\rangle+2\gamma_{m}n_{\text{th}}\langle c_{+}^{\dag}c_{-}\sigma_{z}^{2}\rangle-2\gamma_{m}\langle c_{+}^{\dag}c_{+}^{\dag}c_{+}c_{-}\sigma_{z}^{2}\rangle.
\end{align}
Together with Eq.~(\ref{seq:cplusdagcminus_z}), solving straightforwardly the coupled differential equations (\ref{seq:cplusdagcplusdagcpluscminussigmaz}) and (\ref{seq:cplusdagcplusdagcpluscminussigmaz2}) results in
\begin{align}
\langle c_{+}^{\dag}c_{+}^{\dag}c_{+}c_{-}\sigma_{z}\rangle\left(t\right)=-n_{\text{th}}\left[1-\exp\left(-\gamma_{m}t\right)\right]
\exp\left(-i2Jt\right)\exp\left(-\gamma_{m}t\right),
\end{align}
which, in turn, gives
\begin{equation}
\label{seq:cplusdagcplusdagcpluscminus2} \langle
c_{+}^{\dag}c_{+}^{\dag}c_{+}c_{-}\rangle\left(\tau_{0}\right)=-iY\left(\tau_{0}\right),
\end{equation}
where
\begin{equation}
Y\left(t\right)=n_{\text{th}}\left[1-\exp\left(-\gamma_{m}t\right)\right]\exp\left(-\gamma_{m}t\right).
\end{equation}
In a treatment similar to that used for $\langle c_{+}^{\dag}c_{+}^{\dag}c_{+}c_{-}\rangle$, we obtain
\begin{align}
\label{seq:cplusdagcminusdagcminuscminus}
\langle c_{+}^{\dag}c_{-}^{\dag}c_{-}c_{-}\rangle\left(\tau_{0}\right)&=-iY\left(\tau_{0}\right),\\
\label{seq:cplusdagcplusdagcminuscminus}
\langle c_{+}^{\dag}c_{+}^{\dag}c_{-}c_{-}\rangle\left(\tau_{0}\right)&=0.
\end{align}
Upon combining Eqs.
(\ref{seq:first_interval_cpmdagcpmcdagcpmcpm}),
(\ref{seq:first_intrval_cplusdagcpluscminusdagcminus}),
(\ref{seq:cplusdagcplusdagcpluscminus2}),
(\ref{seq:cplusdagcminusdagcminuscminus}), and
(\ref{seq:cplusdagcplusdagcminuscminus}), this yields, after
inversion back to the localized-mode basis,
\begin{align}
\langle b_{k}^{\dag}b_{k}^{\dag}b_{k}b_{k}\rangle\left(\tau_{0}\right)&=2X\left(\tau_{0}\right),\\
\langle b_{1}^{\dag}b_{1}b_{2}^{\dag}b_{2}\rangle\left(\tau_{0}\right)&=X\left(\tau_{0}\right),\\
\langle b_{1}^{\dag}b_{1}^{\dag}b_{1}b_{2}\rangle\left(\tau_{0}\right)&=\langle b_{1}^{\dag}b_{2}^{\dag}b_{2}b_{2}\rangle\left(\tau_{0}\right)=iY\left(\tau_{0}\right),\\
\langle b_{1}^{\dag}b_{1}^{\dag}b_{2}b_{2}\rangle\left(\tau_{0}\right)&=0.
\end{align}

During the second evolution interval $\tau_{0}<t\leq\tau_{0}+\tau_{1}$, the dynamics of the system is driven by the master equation given in Eq.~(\ref{seq:full-master-equation}), but with $H\left(t\right)$ replaced by $H_{1}$. Thus, as mentioned in Sec.~\ref{sec:Phonon occupations at finite temperature}, the system has a dynamical evolution similar to what has already been discussed with the delocalized-mode basis in the first interval. We follow the same recipe as above and then find
\begin{align}
\label{seq:second_interval_b_1}
\langle b_{k}^{\dag}b_{k}^{\dag}b_{k}b_{k}\rangle\left(\tau_{0}+\tau_{1}\right)&=2X\left(\tau_{0}+\tau_{1}\right),\\
\label{seq:second_interval_b_2}
\langle b_{1}^{\dag}b_{1}b_{2}^{\dag}b_{2}\rangle\left(\tau_{0}+\tau_{1}\right)&=X\left(\tau_{0}+\tau_{1}\right),\\
\label{seq:second_interval_b_3}
\langle b_{1}^{\dag}b_{1}^{\dag}b_{1}b_{2}\rangle\left(\tau_{0}+\tau_{1}\right)&=\langle b_{1}^{\dag}b_{2}^{\dag}b_{2}b_{2}\rangle\left(\tau_{0}+\tau_{1}\right)
=i\exp\left(-iJ\tau_{1}\right)Y\left(\tau_{0}+\tau_{1}\right),\\
\label{seq:second_interval_b_4}
\langle b_{1}^{\dag}b_{1}^{\dag}b_{2}b_{2}\rangle\left(\tau_{0}+\tau_{1}\right)&=0,
\end{align}
at the end of this interval.

For the third evolution interval $\tau_{0}+\tau_{1}<t\leq\tau_{T}$, we return back to the master equation in Eq.~(\ref{seq:delocalized-mode-master-equation}), and also back to the delocalized-mode basis. According to Eqs.~(\ref{seq:second_interval_b_1}), (\ref{seq:second_interval_b_2}), (\ref{seq:second_interval_b_3}), and (\ref{seq:second_interval_b_4}), the evolution at this stage starts from
\begin{align}
\label{seq:second_interval_c_1}
\langle c_{\pm}^{\dag}c_{\pm}^{\dag}c_{\pm}c_{\pm}\rangle\left(\tau_{0}+\tau_{1}\right)&=2X\left(\tau_{0}+\tau_{1}\right)\mp
i2\sin\left(J\tau_{1}\right)Y\left(\tau_{0}+\tau_{1}\right),\\
\label{seq:second_interval_c_2}
\langle c_{+}^{\dag}c_{+}c_{-}^{\dag}c_{-}\rangle\left(\tau_{0}+\tau_{1}\right)&=X\left(\tau_{0}+\tau_{1}\right),\\
\label{seq:second_interval_c_3}
\langle c_{+}^{\dag}c_{+}^{\dag}c_{+}c_{-}\rangle\left(\tau_{0}+\tau_{1}\right)&=\langle c_{+}^{\dag}c_{-}^{\dag}c_{-}c_{-}\rangle\left(\tau_{0}+\tau_{1}\right)=-i\cos\left(J\tau_{1}\right)
Y\left(\tau_{0}+\tau_{1}\right),\\
\langle c_{+}^{\dag}c_{+}^{\dag}c_{+}c_{-}\sigma_{z}^{j}\rangle\left(\tau_{0}+\tau_{1}\right)&=\langle c_{+}^{\dag}c_{-}^{\dag}c_{-}c_{-}\sigma_{z}^{j}\rangle\left(\tau_{0}+\tau_{1}\right)
=i(-1)^{j+1}\sin^{2}\left(\varphi\right)\cos\left(J\tau_{1}\right)
Y\left(\tau_{0}+\tau_{1}\right),\\
\label{seq:second_interval_c_4}
\langle c_{+}^{\dag}c_{+}^{\dag}c_{-}c_{-}\rangle\left(\tau_{0}+\tau_{1}\right)&=\langle c_{+}^{\dag}c_{+}^{\dag}c_{-}c_{-}\sigma_{z}^{j}\rangle\left(\tau_{0}+\tau_{1}\right)=0,
\end{align}
where $j=1,2$. Note that, before this evolution, the spin state has already been transformed from $|D\rangle\rightarrow\cos\left(\varphi\right)|0\rangle+\sin\left(\varphi\right)|D\rangle$ via a single-qubit rotation. Then, by following the same procedure as in the first interval, the last evolution ends with
\begin{align}
\langle c_{\pm}^{\dag}c_{\pm}^{\dag}c_{\pm}c_{\pm}\rangle\left(\tau_{T}\right)&=2X\left(\tau_{T}\right)\pm
2\sin\left(J\tau_{1}\right)Y\left(\tau_{T}\right),\\
\langle c_{+}^{\dag}c_{+}c_{-}^{\dag}c_{-}\rangle\left(\tau_{T}\right)&=X\left(\tau_{T}\right),\\
\langle c_{+}^{\dag}c_{+}^{\dag}c_{+}c_{-}\rangle\left(\tau_{T}\right)&=\langle c_{+}^{\dag}c_{-}^{\dag}c_{-}c_{-}\rangle\left(\tau_{T}\right)=-i\cos\left(J\tau_{1}\right)
\left[\cos^{2}\left(\varphi\right)-i\sin^{2}\left(\varphi\right)\right]Y\left(\tau_{T}\right),\\
\langle c_{+}^{\dag}c_{+}^{\dag}c_{-}c_{-}\rangle\left(\tau_{T}\right)&=0,
\end{align}
and as a result, with
\begin{align}\label{seq:final_b_k}
\langle b_{k}^{\dag}b_{k}^{\dag}b_{k}b_{k}\rangle\left(\tau_{T}\right)&
=2X\left(\tau_{T}\right)+2\left(-1\right)^{j}\sin^{2}\left(\varphi\right)\cos\left(J\tau_{1}\right)Y\left(\tau_{T}\right).
\end{align}
It is seen that on the right-hand side of Eq. (\ref{seq:final_b_k}), the first term arises from the particle behavior of a phonon and the second term arises from its wave behavior.

\begin{figure}[b]
	\centering
	\includegraphics[width=17.0cm]{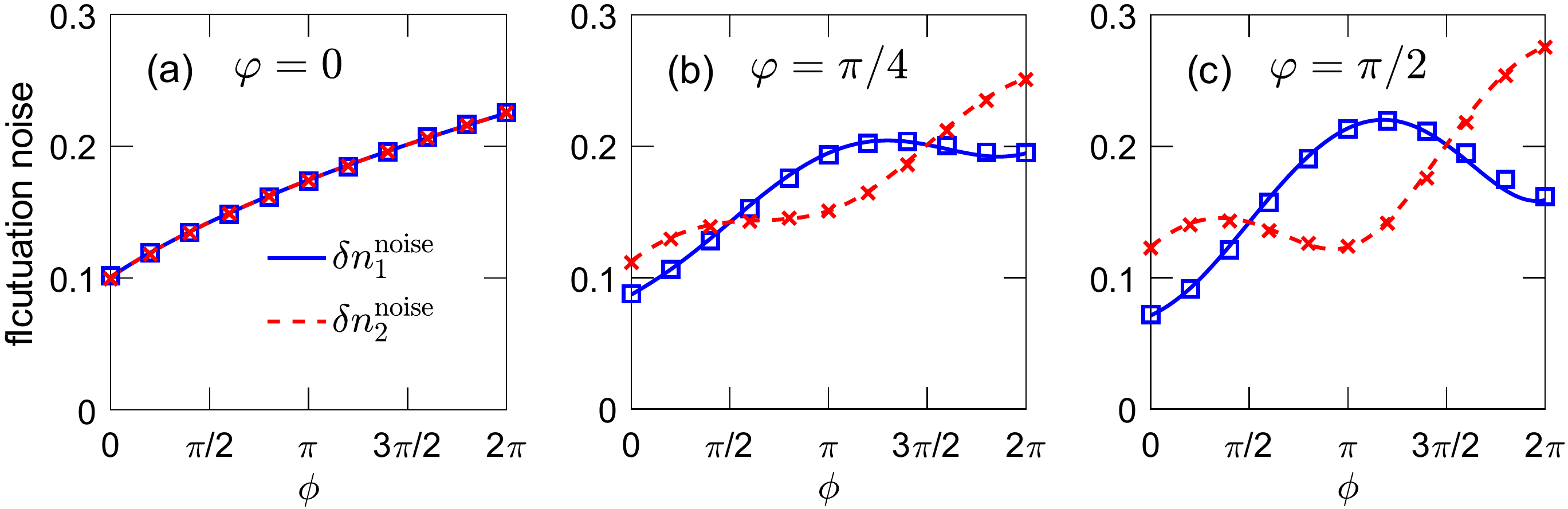}
	\caption{(Color online) Fluctuation noise $\delta n_{1}^{{\rm noise}}$ and $\delta n_{2}^{{\rm noise}}$ as a function of the phase $\phi$. (a) $\varphi=0$, (b) $\pi/4$, and (c) $\pi/2$. Solid and dashed curves are analytical results for $\delta n_{1}^{{\rm noise}}$ and $\delta n_{2}^{{\rm noise}}$, respectively, and symbols correspond to numerical simulations. These analytical and numerical results exhibit an exact agreement. For all plots, in addition to $\gamma_{s}/2\pi=200\gamma_{m}/2\pi=80$~Hz, we have assumed that $g/2\pi=100$~kHz, $\omega_{m}/2\pi=2$~MHz, $\Omega=10\omega_{m}$, and $\Delta_{-}=142\omega_{m}$, resulting in $\omega_{q}\simeq1.5\omega_{m}$ and then $J/2\pi\simeq12$~kHz. We have also assumed that $n_{\text{th}}=100$, corresponding to an environmental temperature of $\simeq 10$~mK.}\label{sfig:s-fluctuation-noise}
\end{figure}

By substituting Eq. (\ref{seq:final_b_k}) into Eq. (\ref{seq:phonon_fluctuation}), the fluctuation $\delta n_{k}$ in the occupation $n_{k}$ is given by
\begin{align}
\left(\delta n_{k}\right)^{2}=&\left(n_{\text{th}}^{2}-2P_{k}n_{\text{th}}-P_{k}^{2}\right)\exp\left(-2\gamma_{m}\tau_{T}\right)\nonumber\\
&-\left(2n_{\text{th}}+1\right)\left(n_{\text{th}}-P_{k}\right)\exp\left(-\gamma_{m}\tau_{T}\right)
+n_{\text{th}}\left(n_{\text{th}}+1\right).
\end{align}
Since $\gamma_{m}\tau_{T}\ll1$, we have
\begin{equation}\label{seq:fluctuation_in_nk}
\left(\delta n_{k}\right)^{2}\simeq\left(\delta n_{k}^{{\rm signal}}\right)^2+\left(\delta n_{k}^{{\rm noise}}\right)^2,
\end{equation}
where
\begin{align}
\left(\delta n_{k}^{{\rm signal}}\right)^{2}&=P_{k}\left(1-P_{k}\right),\\
\label{seq:fluctuation_noise_in_nk}
\left(\delta n_{k}^{\text{noise}}\right)^2&=P_{k}\left(2P_{k}-1\right)\gamma_{m}\tau_{T}
+n_{\text{th}}\gamma_{m}\tau_{T}\left(2P_{k}+1\right).
\end{align}
Here, $\delta n_{k}^{{\rm signal}}$, the quantum fluctuation induced by the Heisenberg uncertainty principle, accounts for the coherent signal, and  $\delta n_{k}^{\text{noise}}$ represents the fluctuation noise, including the vacuum (the first term) and thermal (the second term) fluctuations. To confirm the predictions of Eq.~(\ref{seq:fluctuation_in_nk}), we perform numerics, as shown in Fig.~\ref{sfig:s-fluctuation-noise}. Specifically, we plot the fluctuation noises $\delta n_{1}^{\text{noise}}$ and $\delta n_{2}^{\text{noise}}$ versus the relative phase $\phi$. The analytical expression is in excellent agreement with our numerical simulations. Furthermore, the respective CNT signal-to-noise ratios can be defined as
\begin{figure}[t]
	\centering
	\includegraphics[width=10.0cm]{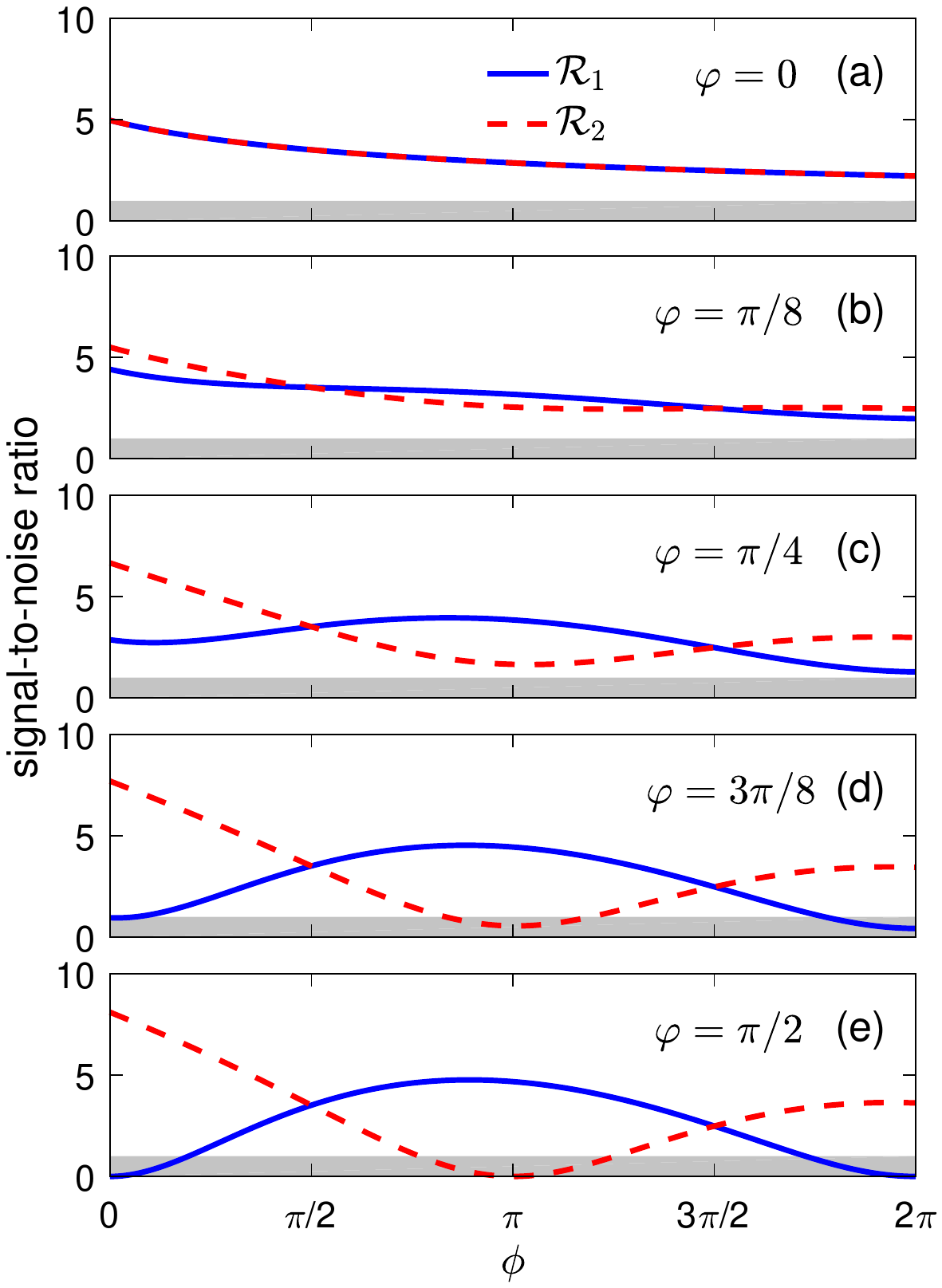}
	\caption{(Color online) Signal-to-noise ratios $\mathcal{R}_{1}$ and $\mathcal{R}_{2}$. (a) $\varphi=0$, (b) $\pi/8$, (c) $\pi/4$, (d) $3\pi/8$, and (e) $\pi/2$. The solid curves show $\mathcal{R}_{1}$, while the dashed curves show $\mathcal{R}_{2}$. The gray shaded area represents the region, where the signal cannot be resolved. For all plots, all other parameters have been set to be the same as in Fig. \ref{sfig:s-fluctuation-noise}.}\label{sfig:s-signal-to-noise-ratio-n1-n2}
\end{figure}
\begin{equation}\label{seq:signal-to-noise_ratio_of_nk}
\mathcal{R}_{k}=\frac{P_{k}}{\delta n_{k}^{{\rm noise}}}.
\end{equation}
Note that, here, we did not use $\delta n_{k}$ to define $\mathcal{R}_{k}$ because $\delta n_{k}^{{\rm signal}}$ in $\delta n_{k}$ results from quantum fluctuations of the desired signal, as mentioned previously; and therefore this is not the environmental noise. In order to resolve a signal from the fluctuation noise, the ratio $\mathcal{R}_{k}$ is required to be $\mathcal{R}_{k}>1$. However, Eq. (\ref{seq:signal-to-noise_ratio_of_nk}) demonstrates that this criterion is not always met for all values of $P_{k}$, in particular, at finite temperatures. For example, $P_{k}=0$ leads directly to $\mathcal{R}_{k}=0$. To address this problem, we now consider the total fluctuation noise,
\begin{equation}
\mathcal{S}^{2}=\left(\delta n_{1}^{\text{noise}}\right)^{2}+\left(\delta n_{2}^{\text{noise}}\right)^{2}.
\end{equation}
We further assume that
\begin{equation}\label{seq:nT_upper_bound}
\mathcal{S}^{2}<P_{1}^{2}+P_{2}^{2}.
\end{equation}
Under this assumption, if $\mathcal{R}_{k}<1$, then $\mathcal{R}_{3-k}>1$ for $k=1,2$; otherwise $\mathcal{R}_{1}>1$, $\mathcal{R}_{2}>1$. This means that at least one of the signals, $P_{1}$ or $P_{2}$, is resolved for each measurement. Because the coherent phonon number equal to $1$ is conserved, and therefore the signals in the two CNTs are complementary, the unresolved signal can be completely deduced from the resolved one. Thus, the criterion in Eq.~(\ref{seq:nT_upper_bound}) ensures that the morphing behavior between wave and particle can be observed from the environment-induced fluctuation noise. In fact, for any value of $P_{k}$, the total noise $\mathcal{S}$ is limited by an upper bound,
\begin{equation}
\mathcal{S}<\mathcal{B}\equiv\sqrt{\gamma_{m}\tau_{T}^{{\rm max}}+4n_{{\rm th}}\gamma_{m}\tau_{T}^{{\rm max}}},
\end{equation}
which is independent of $P_{k}$. Meanwhile, $\sqrt{P^{2}_{1}+P^{2}_{2}}$ is also limited by a lower bound $\sqrt{2}/2$. Thereby, in order to meet the criterion given in Eq.~(\ref{seq:nT_upper_bound}), it is required that
\begin{equation}\label{seq:criterion}
\mathcal{B}<\frac{\sqrt{2}}{2}.
\end{equation}
Based on this condition, we can define a signal visibility
\begin{equation}
\mathcal{R}=\frac{\sqrt{2}}{2\mathcal{B}},
\end{equation}
in analogy to $R_{k}$. When $\mathcal{R}>1$, the morphing between wave and particle can be observed, and cannot otherwise. This, in turn, leads to an upper bound on the equilibrium phonon occupation,
\begin{equation}
n_{{\rm th}}<\frac{1-2\gamma_{m}\tau_{T}^{{\rm max}}}{8\gamma_{m}\tau_{T}^{{\rm max}}},
\end{equation}
and therefore an upper bound on the temperature,
\begin{equation}
T<\frac{\hbar\omega_{m}}{k_{B}\ln\left[\left(1+6\gamma_{m}\tau_{T}^{{\rm max}}\right)/\left(1-2\gamma_{m}\tau_{T}^{{\rm max}}\right)\right]}.
\end{equation}
Because $\tau_{T}^{{\rm max}}\simeq5\pi/2J$, the critical temperature is
\begin{equation}\label{seq:Tc}
T_{\rm c}=\frac{\hbar\omega_{m}}{k_{B}\ln\left[\left(1+15\pi\gamma_{m}/J\right)/\left(1-5\pi\gamma_{m}/J\right)\right]}.
\end{equation}
In Fig. \ref{sfig:s-signal-to-noise-ratio-n1-n2} we plot the signal-to-noise ratios $\mathcal{R}_{1}$ and $\mathcal{R}_{2}$ at the temperature $T\simeq10$~mK. We find that almost all signals can be resolved, and also, as expected, find that when the signal in one CNT is unresolved, the signal in the other CNT is resolved. In fact, the upper bound $\mathcal{B}$ is the fluctuation noise in the total phonon occupation $\langle b_{1}^{\dag}b_{1}+b_{2}^{\dag}b_{2}\rangle$ at time $\tau_{T}$. The criterion $\mathcal{R}>1$ heralds that to resolve the morphing behavior, the fluctuation noise in $\langle b_{1}^{\dag}b_{1}+b_{2}^{\dag}b_{2}\rangle\left(\tau_{T}\right)$ is required to be smaller than $\sqrt{2}/2$.

\section{Numerical simulations}
\label{sec:numerical simulations}

\begin{figure}[tbph]
	\centering
	\includegraphics[width=17.6cm]{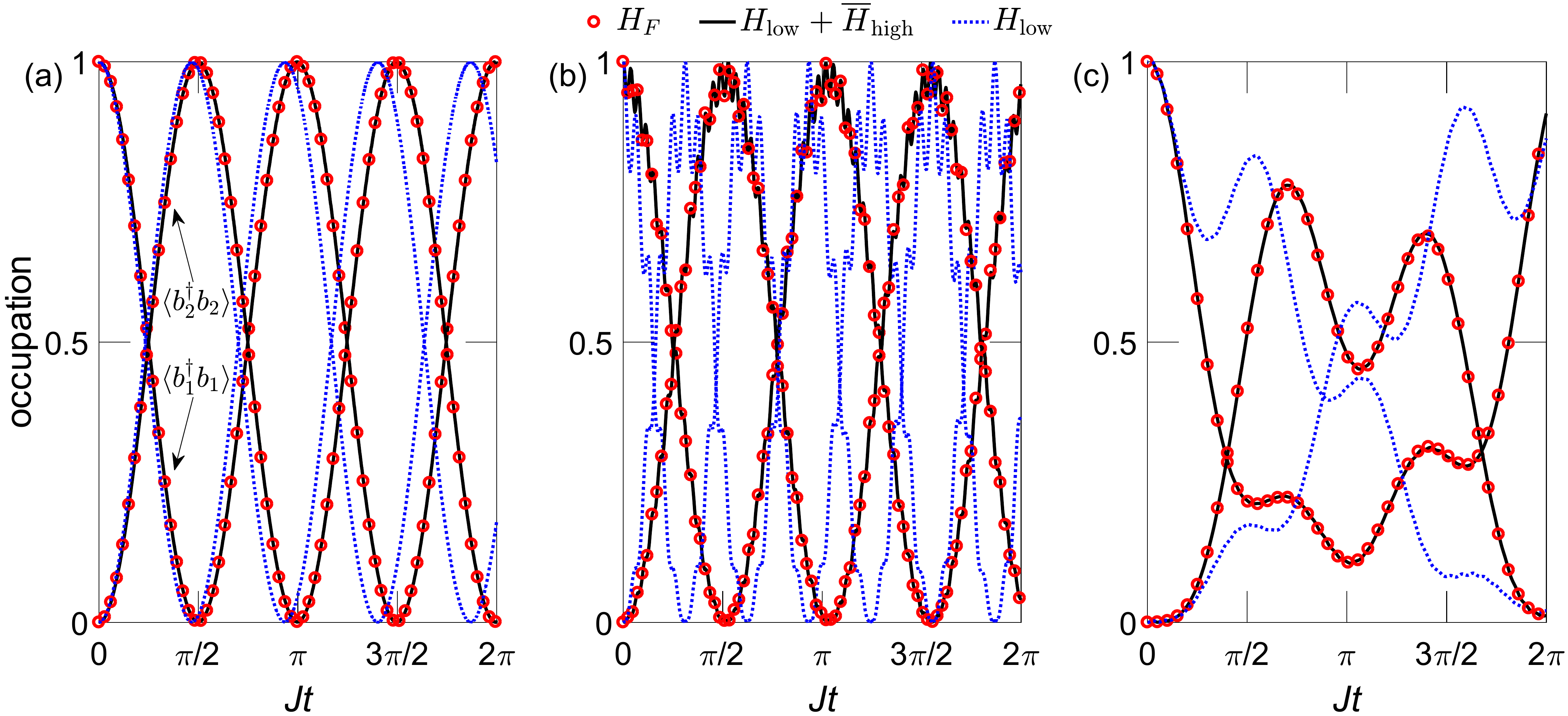}
	\caption{(Color online) Unitary evolution of the CNT phonon occupations, $\langle b_{1}^{\dag}b_{1}\rangle$ and $\langle b_{2}^{\dag}b_{2}\rangle$, for (a) $\Delta_{-}=10\Omega$, (b) $25\Omega$, and (c) $35\Omega$. The symbols, solid, and dotted curves are obtained, respectively, from $H_{F}$, $H_{\text{low}}+H_{\text{high}}$, and $H_{\text{low}}$. For all plots here we have assumed that $\omega_{m}/2\pi=2$~MHz, $\Omega=15\omega_{m}$, $\omega_{0}=D-\Delta_{-}$, $\Delta_{+}=D+\omega_{0}$, $\Delta=\Delta_{-}+3\Omega^{2}/\Delta_{+}$, $\omega_{q}=2\Omega^{2}/\Delta+2\Omega^{2}/\Delta_{+}$,
		and $J=2\omega_{q}g^{2}/\left(\omega_{q}^{2}-\omega_{m}^{2}\right)$, with a symmetric coupling strength $g/2\pi=100$~kHz and an initial state $|\Psi\rangle_{i}=\left(b_{1}^{\dag}\otimes\mathcal{I}_{2}
		|\text{vac}\rangle\right)\otimes|D\rangle$.}
	\label{sfig:s-confirm-time-indenpendent-full-Hamiltonian}
\end{figure}

In order to confirm our analytical results, we need to numerically simulate the dynamics with the full master equation given by
\begin{align}\label{seq:exact-full-master-equation}
\dot{\rho}\left(t\right)=&\frac{i}{\hbar}\left[\rho\left(t\right),H_{F}\right]
-\frac{\gamma_{s}}{2}\mathcal{L}\left(\sigma_{z}^{\prime}\right)\rho\left(t\right)\nonumber\\
&-\frac{\gamma_{m}}{2}n_{\text{th}}\sum_{k=1,2}\mathcal{L}\left(b_{k}^{\dag}\right)\rho\left(t\right)
-\frac{\gamma_{m}}{2}\left(n_{\text{th}}+1\right)\sum_{k=1,2}\mathcal{L}\left(b_{k}\right)\rho\left(t\right),
\end{align}
where $\sigma_{z}^{\prime}=|D\rangle\langle D|-|0\rangle\langle 0|$, and $H_{F}$ is the full Hamiltonian of Eq. (\ref{seq:original-full-Hamiltonian}). Here, we use the Python framework QuTiP~\cite{johansson2012qutip,johansson2013qutip2} to set up this problem. However, the full Hamiltonian is time-dependent, and it takes a long time to integrate the corresponding Schr\"{o}dinger equation or the master equation, in particular, for our case, where all quantum gates result from the deterministic time evolution of the system. Thus, in our numerical simulations, we replace $H_{F}$ with $H_{\text{low}}+\overline{H}_{\text{high}}$, as in Eq.~(\ref{seq:time-independent-full-Hamiltonian}). This is a reasonable replacement because in our proposal $\Omega$ (tens of~MHz) is required to be much smaller than $\Delta^{\prime}$ (up to $\sim$~GHz).
In Fig.~\ref{sfig:s-confirm-time-indenpendent-full-Hamiltonian},
we plot the unitary evolution of the phonon occupations, $\langle b_{1}^{\dag}b_{1}\rangle$ and $\langle b_{2}^{\dag}b_{2}\rangle$, of the CNTs. Symbols are the exact results from the full Hamiltonian $H_{F}$ and solid curves are given by the approximate Hamiltonian $H_{\text{low}}+\overline{H}_{\text{high}}$. We find an excellent agreement for a very long evolution time, and thus $H_{F}$ can be very well approximated by $H_{\text{low}}+\overline{H}_{\text{high}}$. For additional comparison, we also plot the phonon occupation evolution driven only by the low-frequency component $H_{\text{low}}$, corresponding to dotted curves. As seen in Fig.~\ref{sfig:s-confirm-time-indenpendent-full-Hamiltonian}, owing to the error accumulation, the dynamics of $H_{\text{low}}$ deviates largely from the full dynamics of $H_{F}$, even within one oscillation cycle. With the above replacement, we obtain the numerical simulations plotted in Fig.~2 of the article, and also in Fig.~\ref{sfig:s-fluctuation-noise} of the Supplemental Material.

\newpage

\bibliographystyle{naturemag}

\end{document}